%% file: main.tex
\documentclass[article]{IEEEtran}
\usepackage{amsmath}
\usepackage{tikz}
\usepackage{cite}
\usepackage{circuitikz}
 \usepackage{tkz-euclide}
 \usetikzlibrary{angles}
 \usetikzlibrary{positioning}
\usepackage{tikz-3dplot}
\usepackage[utf8]{inputenc}
\usepackage{color}
\usepackage{soul}
\usepackage{amssymb}

\newcommand{\M}[1]{\mathbf{#1}}
\newcommand{\T}[1]{\mathrm{#1}}
\newcommand{\V}[1]{\boldsymbol{#1}}

\newcommand{\bs}[1]{\boldsymbol{\V{#1}}}
\renewcommand{\u}[1]{\boldsymbol{\hat{#1}}}
\usetikzlibrary{shapes.symbols}
\newcommand{\Jh}{\hat{\T{J}}}

\newcommand{\Hh}{\hat{\T{H}}^{(2)}}
\newcommand{\Hhp}{\hat{\T{H}}^{(2)\prime}}
\newcommand{\rbkind}{\chi}
\newcommand{\Xopchar}{P}
\newcommand{\Yopchar}{Q}
\newcommand{\Xop}{\mathcal{\Xopchar}}
\newcommand{\Xmat}{\mathbf{\Xopchar}}
\newcommand{\Yop}{\mathcal{\Yopchar}}
\newcommand{\Ymat}{\mathbf{\Yopchar}}
\newcounter{tempEQCounter}

\title{Scattering Properties of Spherical Time-Varying Conductive Shells}
\author{Kurt Schab, \IEEEmembership{Member, IEEE}, Bradley Shirley, \IEEEmembership{Student Member, IEEE}, and K.C. Kerby-Patel \IEEEmembership{Senior Member, IEEE}
\thanks{Manuscript received  \today; revised \today.}
\thanks{K. Schab and B. Shirley are with the Department of Electrical and Computer Engineering, Santa Clara University, Santa Clara, CA, USA (e-mail: kschab@scu.edu).}
\thanks{K. C. Kerby-Patel is with the Engineering Department, University of Massachusetts Boston, Boston, MA USA (e-mail: kc.kerby-patel@umb.edu).}}
\begin{document}

\maketitle
\setcounter{page}{1}

\maketitle

\begin{abstract}
    Harmonic generation in the scattered fields produced by a dielectric sphere coated with a time-varying conductive shell is studied using a Mie theory approach hybridized with conversion matrix methods.  Analytic results are derived for plane wave incidence as well as general external excitation using transition matrix techniques.  An equivalent transmission line approach is also discussed.  Numerical examples validate the derived expressions through comparison with purely numerical methods and convergence characteristics are explored.  Several additional examples illustrate unique trends in far- and near-field scattering.  
\end{abstract}


\section{Introduction}

\IEEEPARstart{T}{ime-varying} electromagnetic devices exhibit frequency conversion, i.e., the generation of spectral content not present in an applied excitation.  Historically, time-varying elements have been commonplace within circuit systems used to perform up- or down-conversion. More recently, spatiotemporally (space-time) modulated materials have been applied to bring about similar frequency conversion phenomena in distributed systems.  The analytic study of space-time materials has, so far, been largely driven by methods involving homogeneous media\cite{hayrapetyan2016electromagnetic,caloz2019spacetime,koutserimpas2020electromagnetic}, the polarizability of single particles \cite{mirmoosa2020dipole},~
guided wave systems \cite{hadad2020soft}, planar structures \cite{morgenthaler1958velocity,holberg1966parametric,Fante1973OnTP,harfoush1991scattering,wu2019serrodyne,salary2018electrically,Inampudi:19,pantazopoulos2019layered,taravati2020space,li2021temporal}, or systems with inherent cylindrical symmetry \cite{Wright_2010,salary2018}.

Many of these analytic studies, explicitly or implicitly, apply conversion matrix techniques \cite{maas2003nonlinear} commonly used in the study of time-varying circuits to analyze harmonic generation. Within this framework, linear, time-varying components or materials act as coupling elements between otherwise decoupled linear, time-invariant physical problems at multiple frequencies.  Recently, this approach was integrated with the method of moments as an alternative to the finite difference time domain (FDTD) method for the analysis of arbitrary time-varying electromagnetic systems \cite{palmer2019investigation,bass2021conversion}.  Despite the flexibility of FDTD and other numerical methods as computational tools for studying time-varying systems, analytic or quasi-analytic methods remain attractive due to their low computational cost and their ability to accurately model many problems based on canonical geometries.

In this work, we develop an analytic solution for the fields scattered by a dielectric sphere coated with a thin shell of time-varying conductivity.  This study is motivated by the implementation of time-varying conductivity through dense networks of lumped time-varying components (e.g., switches as in \cite{wu2019serrodyne,taravati2020space}), the modulation of inherently surface-based materials (e.g., carbon nanostructures \cite{salary2018electrically}),  or temporal modulation of plasma ionization \cite{singletary2021}.~ The solution leverages aspects of both classical Mie scattering theory \cite{jin2011theory} as well as conversion matrix \cite{maas2003nonlinear} methodologies.  It shares many features with the solution of analogous problems involving static impedance boundary conditions \cite{wait1965calculations, sihvola2018resonances}, time-varying planar and cylindrical structures \cite{censor2004non,Wright_2010,salary2018}, time-varying dielectric spheres \cite{stefanou2021light}, and spheres constructed from non-linear media \cite{dewitz1996theory,pavlyukh2004nonlinear}; though the formulation and examples studied here are unique to the spherical shell configuration.  

In an effort to establish connections between the time-varying shell geometry and other scattering problems (both simpler and more complex), we carry out this analysis by several means, each with a unique engineering perspective, generality, and adaptability to other problems.  We begin by approaching the problem via classical Mie theory for the special case of plane wave incidence using explicit forms of all field components.  We then generalize the approach to arbitrary external excitation to arrive at a transition (T-) matrix formulation compatible with general external excitation and multi-scatterer systems \cite{waterman1965matrix}.  Finally, we adapt the T-matrix formulation into a transmission line problem giving rise to a conversion matrix form of the reflection coefficient observed at an impedance interface.  Several examples are then discussed and validated against numerical methods, namely conversion matrix method of moments (CMMoM).  Trends in far-field scattering and near-field focusing behavior are explored through supplementary examples.

\section{Scattering due to a plane wave}
\label{sec:pw}
In this section, we derive a quasi-analytic solution for the field scattered by a sphere with time-varying properties.  Though the solution itself is made more complex by the time-varying nature of the system, the overall strategy remains the same as that used in static boundary value problems.  We begin by expanding unknown internal and scattered fields in terms of known basis functions, then proceed to determine unknown expansion coefficients through the application of appropriate boundary conditions.

\begin{figure}
    \centering
   \input{figures/fig-01-sphere}
    \caption{A sphere of radius $a$ contains a static, non-dispersive material with permittivity and permeability $\varepsilon_\T{d} = \varepsilon_\T{r}\varepsilon_0$ and $\mu_\T{d} = \mu_\T{r}\mu_0$, respectively.  The surface of the sphere is coated with a time-varying surface conductivity $\sigma(t)$.}
    \label{fig:schem}
\end{figure}
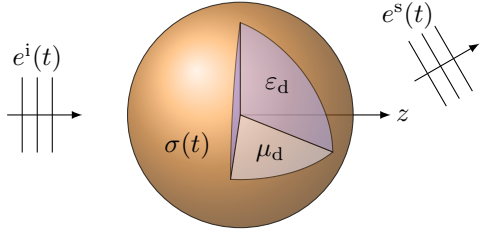

Throughout this work we consider a sphere of radius $a$ containing a core of homogeneous material with permittivity $\varepsilon_\T{d} = \varepsilon_\T{r}\varepsilon_0$ and permeability $\mu_\T{d} = \mu_\T{r}\mu_0$.  The surface of this sphere has a time-varying surface conductivity $\sigma(t)$ with Fourier representation $\hat{\sigma}(\omega)$, as shown in Fig.~\ref{fig:schem}.  Here we have assumed that the causal response (relaxation) time of the material is very short with respect to any signal variations within the system such that tangential electric fields $e_\T{t}(t)$ and surface currents $j(t)$ are related instantaneously through the constitutive relation 
\begin{equation}
    j(t) = \int_0^\infty \sigma(t,\tau)e_\T{t}(t-\tau)\T{d}\tau \approx \sigma(t)e_\T{t}(t).
\end{equation}
Note that while this approximation of $\sigma(t,\tau)\approx \sigma(t)\delta(\tau)$ is non-causal, it is the standard approach to modeling both static and dynamic components with negligible dispersive effects over the frequency range of interest~\cite[\S 3.4.1]{maas2003nonlinear}. For detailed discussion of modeling dispersive effects in dynamic lumped components, see~\cite{jayathurathnage2021time}.  Those methods can be applied to the process utilized in this work without considerable difficulty, but here we consider non-causal ``instantaneous'' time-varying media for simplified notation and to focus on the key conceptual steps in the presented approach for time-varying analysis.  This approximation is naturally most accurate when all time scales\footnote{``All time scales'' here refers to characteristic times and periods of incident fields, scattered harmonic fields, and all time-variations of the surface conductivity.} are very long relative to any material relaxation times~$\tau_\T{r}$ manifested by dispersion, e.g., at HF frequencies (3-30 MHz, $T\sim30$~ns) and conductivities within a few orders of magnitude of copper ($\tau_\T{r}\sim10^{-19}$~s).

We begin by considering a situation where the sphere is illuminated by a multi-frequency incident plane wave $\V{e}_\T{i}(t)$ and $\V{h}_\T{i}(t)$ with Fourier components of the form
\begin{equation}
    \V{E}^\T{i}(\omega) = \u{x}E(\omega)\T{e}^{-\T{j}kz},\quad
    \V{H}^\T{i}(\omega) = \u{y}\eta^{-1}E(\omega)\T{e}^{-\T{j}kz},
\end{equation}
where $k$ is the free space wavenumber at frequency $\omega$ and $\eta$ is the free space wave impedance.  At each frequency this wave may be expanded into spherical harmonics, leading to the tangential field components \cite[\S 7.4.3]{jin2011theory}
\begin{multline} \label{eq:pwexp-et}
    E_\theta^\T{i}(\omega) = - \frac{E_0 g(\omega)\cos\phi}{kr}\sum_{n=1}^{\infty} \T{j}^{-n}\frac{2n+1}{n(n+1)}\\\times\biggl[\T{j}\Jh'_n(kr)\frac{\T{d}\T{P}_n^1(\cos\theta)}{\T{d}\theta}+\Jh_n(kr)\frac{\T{P}_n^1(\cos\theta)}{\sin\theta}\biggr]
\end{multline}
\begin{multline}
    E_\phi^\T{i}(\omega) = \frac{E_0 g(\omega)\sin\phi}{kr}\sum_{n=1}^{\infty} \T{j}^{-n}\frac{2n+1}{n(n+1)}\\\times\left[\T{j}\Jh'_n(kr)\frac{\T{P}_n^1(\cos\theta)}{\sin\theta}+\Jh_n(kr)\frac{\T{d}\T{P}_n^1(\cos\theta)}{\T{d}\theta}\right]
\end{multline}
\begin{multline}
    H_\theta^\T{i}(\omega) = - \frac{E_0 g(\omega)\sin\phi}{\eta kr}\sum_{n=1}^{\infty} \T{j}^{-n}\frac{2n+1}{n(n+1)}\\\times\left[\Jh_n(kr)\frac{\T{P}_n^1(\cos\theta)}{\sin\theta}+\T{j}\Jh'_n(kr)\frac{\T{d}\T{P}_n^1(\cos\theta)}{\T{d}\theta}\right]
\end{multline}
\begin{multline}\label{eq:pwexp-hp}
    H_\phi^\T{i}(\omega) = - \frac{E_0 g(\omega)\cos\phi}{\eta kr}\sum_{n=1}^{\infty} \T{j}^{-n}\frac{2n+1}{n(n+1)}\\\times\left[\Jh_n(kr)\frac{\T{d}\T{P}_n^1(\cos\theta)}{\T{d}\theta}+\T{j}\Jh'_n(kr)\frac{\T{P}_n^1(\cos\theta)}{\sin\theta}\right]
\end{multline}
where $\Jh_n$ are spherical Riccati-Bessel functions and $\T{P}_n^1$ are associated Legendre polynomials.  Here $E_0$ represents a reference field magnitude while the unitless factor $g(\omega)$ indicates spectral weighting.

\input{equations/g-eq}

We may also expand the unknown scattered and internal fields at each frequency into similar spherical harmonics, matching the form of terms in the spherical wave expansion of the incident plane wave.  For the scattered field, this expansion reads
\begin{multline}
    E_\theta^\T{s}(\omega) = -\frac{E_0\cos\phi}{kr}\sum_{n=1}^\infty\\\left[a_n\T{j}\Hhp_n(kr)\frac{\T{d}\T{P}_n^1(\cos\theta)}{\T{d}\theta}+b_n\Hh_n(kr)\frac{\T{P}_n^1(\cos\theta)}{\sin\theta}\right]
    \label{eq:E-sc-t}
\end{multline}
\begin{multline}
    E_\phi^\T{s}(\omega) = \frac{E_0\sin\phi}{kr}\sum_{n=1}^\infty\\\left[a_n\T{j}\Hhp_n(kr)\frac{\T{P}_n^1(\cos\theta)}{\sin\theta}+b_n\Hh_n(kr)\frac{\T{d}\T{P}_n^1(\cos\theta)}{\T{d}\theta}\right]
    \label{eq:E-sc-p}
\end{multline}
\begin{multline}
    H_\theta^\T{s}(\omega) = -\frac{E_0\sin\phi}{\eta kr}\sum_{n=1}^\infty\\\left[a_n\Hh_n(kr)\frac{\T{P}_n^1(\cos\theta)}{\sin\theta}+b_n\T{j}\Hhp_n(kr)\frac{\T{d}\T{P}_n^1(\cos\theta)}{\T{d}\theta}\right]
    \label{eq:H-sc-t}
\end{multline}
\begin{multline}
    H_\phi^\T{s}(\omega) = -\frac{E_0\cos\phi}{\eta kr}\sum_{n=1}^\infty\\\left[a_n\Hh_n(kr)\frac{\T{d}\T{P}_n^1(\cos\theta)}{\T{d}\theta}+b_n\T{j}\Hhp_n(kr)\frac{\T{P}_n^1(\cos\theta)}{\sin\theta}\right]
    \label{eq:H-sc-p}
\end{multline}
and for the internal field we have
\begin{multline}
    E_\theta^\T{d}(\omega) =  -\frac{E_0\cos\phi}{k_\T{d}r}\sum_{n=1}^\infty\\\left[c_n\T{j}\Jh_n'(k_\T{d}r)\frac{\T{d}\T{P}_n^1(\cos\theta)}{\T{d}\theta}+d_n\Jh_n(k_\T{d}r)\frac{\T{P}_n^1(\cos\theta)}{\sin\theta}\right]
\end{multline}
\begin{multline}
    E_\phi^\T{d}(\omega) = \frac{E_0\sin\phi}{k_\T{d}r}\sum_{n=1}^\infty\\\left[c_n\T{j}\Jh_n'(k_\T{d}r)\frac{\T{P}_n^1(\cos\theta)}{\sin\theta}+d_n\Jh_n(k_\T{d}r)\frac{\T{d}\T{P}_n^1(\cos\theta)}{\T{d}\theta}\right]
\end{multline}
\begin{multline}
    H_\theta^\T{d}(\omega) = -\frac{E_0\sin\phi}{\eta_\T{d}k_\T{d}r}\sum_{n=1}^\infty\\\left[c_n\Jh_n(k_\T{d}r)\frac{\T{P}^1_n(\cos\theta)}{\sin\theta} + d_n\T{j}\Jh_n'(k_\T{d}r)\frac{\T{d}\T{P}_n^1(\cos\theta)}{\T{d}\theta}\right]
\end{multline}
\begin{multline}
    H_\phi^\T{d}(\omega) =  -\frac{E_0\cos\phi}{\eta_\T{d}k_\T{d}r}\sum_{n=1}^\infty\\\left[c_n\Jh_n(k_\T{d}r)\frac{\T{d}\T{P}_n^1(\cos\theta)}{\T{d}\theta} + d_n\T{j}\Jh_n'(k_\T{d}r)\frac{\T{P}^1_n(\cos\theta)}{\sin\theta}\right].
\end{multline}
In the above expressions, $k_\T{d}$ and $\eta_\T{d}$ denote the wavenumber and characteristic impedance of the core material, respectively.  The coefficients $\{a_n\}$ and $\{c_n\}$ are associated with transverse magnetic (TM) modes, while $\{b_n\}$ and $\{d_n\}$ correspond to transverse electric (TE) modes. 

Tangential field continuity at the boundary of the sphere gives rise to four boundary conditions that must hold for all times $t$, specifically
\begin{equation}
    e_\theta^\T{i}(t)+e_\theta^\T{s}(t)=
    e_\theta^\T{d}(t)
    \label{eq:e-theta-cont}
\end{equation}
\begin{equation}
    e_\phi^\T{i}(t)+e_\phi^\T{s}(t) = 
    e_\phi^\T{d} (t)
    \label{eq:e-phi-cont}
\end{equation}
\begin{equation}
    h_\theta^\T{i}(t)+h_\theta^\T{s}(t) = 
    h_\theta^\T{d}(t) + \sigma(t)e_\phi^\T{d}(t)
    \label{eq:h-theta-cont}
\end{equation}
\begin{equation}
    h_\phi^\T{i}(t)+h_\phi^\T{s}(t) = 
    h_\phi^\T{d}(t) - \sigma(t)e_\theta^\T{d}(t).
    \label{eq:h-phi-cont}
\end{equation}
Within the magnetic field boundary conditions \eqref{eq:h-theta-cont} and \eqref{eq:h-phi-cont}, terms of the form $\sigma(t)e_{x}^\T{d}(t)$ represent a surface current density. 
Temporal phase matching gives analogous conditions on each Fourier component of the fields.  Applying the series forms of incident, internal, and scattered fields in the frequency domain and matching terms of like tangential dependence, the first pair of boundary conditions \eqref{eq:e-theta-cont} and \eqref{eq:e-phi-cont} gives
\begin{equation}
    g_n\Jh'_n(ka) + a_n\Hhp_n(ka) = \frac{k}{k_\T{d}}c_n\Jh_n'(k_\T{d}a)
    \label{eq:bc-e-tm}
\end{equation}
\begin{equation}
    g_n\Jh_n(ka) + b_n\Hh_n(ka) =\frac{k}{k_\T{d}}d_n\Jh_n(k_\T{d}a)
\end{equation}
where 
\begin{equation}
    g_n = g(\omega)\T{j}^{-n}\frac{2n+1}{n(n+1)}
\end{equation}
and the frequency dependence of $g_n$, $a_n$, $b_n$, $c_n$, and $d_n$ has been suppressed.  Since multiplication in the time domain results in convolution in the frequency domain, the second set of boundary conditions in \eqref{eq:h-theta-cont} and \eqref{eq:h-phi-cont} is more complex, reading 
\begin{multline}
    g_n\Jh_n(ka)+a_n\Hh_n(ka) = \frac{\mu}{\mu_\T{d}}c_n \Jh_n(k_\T{d}a) \\- \T{j} k\eta\int_{-\infty}^\infty k_\T{d}^{\circ-1} \hat{\sigma}(\omega-\omega^\circ)c_n(\omega^\circ)\Jh_n'(k^\circ_\T{d}a) \T{d}\omega^\circ,
    \label{eq:bc-h-tm}
\end{multline}
\begin{multline}
    g_n\Jh_n'(ka)+b_n\Hhp_n(ka) = \frac{\mu}{\mu_\T{d}}d_n \Jh_n'(k_\T{d}a) \\+ \T{j} k\eta\int_{-\infty}^\infty k_\T{d}^{\circ-1} \hat{\sigma}(\omega-\omega^\circ)d_n(\omega^\circ)\Jh_n(k^\circ_\T{d}a) \T{d}\omega^\circ.
\end{multline}
Here $^\circ$ is used to mark integration variables as $'$ is already used to denote differentiation of a function with respect to its argument.  Considering the functions $g_n$, $a_n$, $b_n$, $c_n$, and $d_n$ described by a basis of monochromatic signals, we may interpret all terms in the above boundary conditions as diagonal operators, with the exception of the convolution terms which naturally give rise to cross frequency coupling.  As expected, we see that the TE and TM excitation coefficients are fully decoupled.

We first consider the system of equations governing the TM coefficients.  Solving \eqref{eq:bc-e-tm} for the external coefficients $a_n$ gives
\begin{equation}
     a_n = \frac{1}{\Hhp_n(ka)}\left[\frac{k}{k_\T{d}}c_n\Jh_n'(k_\T{d}a) - g_n\Jh_n'(ka)\right].
    \label{eq:an-cn}
\end{equation}
Using this expression to eliminate the external coefficients from \eqref{eq:bc-h-tm} and collecting terms leads to \eqref{eq:g-c-tm-long}\stepcounter{equation}, where the Riccati-Bessel function Wronskian identity~\cite{abramowitz1948handbook}
\begin{equation}
    \Jh_n(z)\Hhp_n(z) - \Jh_n'(z)\Hh_n(z) = -\T{j}
    \label{eq:h2-wronsk}
\end{equation}
has been used.  Defining the operators
\begin{multline}
    \mathcal{A}_n x = \T{j}^{-1}\mu_\T{r}^{-1}\sqrt{\varepsilon_\T{r}}^{-1}\\\times\left[\sqrt{\mu_\T{r}}\Hh_n(ka)\Jh_n'(k_\T{d}a) -\sqrt{\varepsilon_\T{r}}\Hhp_n(ka)\Jh_n(k_\T{d}a)\right]x
\end{multline}
and
\begin{multline}
    \mathcal{B}_n x = \omega\sqrt{\varepsilon_\T{r}\mu_\T{r}}^{-1}\Hhp_n(ka)\\\times\int_{-\infty}^\infty \omega^{\circ-1}\eta\hat{\sigma}(\omega-\omega^\circ)x(\omega^\circ)\Jh_n'(k^\circ_\T{d}a) \T{d}\omega^\circ
\end{multline}
allows for a condensed form of \eqref{eq:g-c-tm-long} to be written as
\begin{equation}
    g_n = \left(\mathcal{A}_n+\mathcal{B}_n\right)c_n.
    \label{eq:gn-tm-cont}
\end{equation}
Again, with respect to a basis of monochromatic signals, the operator $\mathcal{A}_n$ is always diagonal, whereas the operator $\mathcal{B}_n$ associated with the surface conductivity is, in general, not diagonal.  From this reduced operator expression, we immediately recognize simplifications arising from several important special cases.  First, when the surface conductivity is identically zero the operator $\mathcal{B}_n$ vanishes, yielding the classical TM scattering coefficients for a dielectric sphere.  Similarly, if the surface conductivity is purely static (not varying in time) the operator $\mathcal{B}_n$ becomes diagonal~\cite{wait1965calculations}. Additionally, if the inner medium has the same material properties as the external medium (i.e., $\varepsilon_\T{r} = \mu_\T{r} = 1$), then $\mathcal{A}_n$ becomes the identity operator.  Two special cases reduce the above system to one of a PEC sphere, either $\sigma(t)\rightarrow\infty$ for all time or  $\varepsilon_\T{r}\rightarrow -\T{j}\infty$.

Repeating the above analysis for TE modes we find the external coefficients to be
\begin{equation}
    b_n = \frac{1}{\Hh_n(ka)}\left[\frac{k}{k_\T{d}}d_n\Jh_n(k_\T{d}a) - g_n\Jh_n(ka)\right]
    \label{eq:bn-dn}
\end{equation}
and the internal coefficients given by
\begin{equation}
    g_n = (\mathcal{C}_n + \mathcal{D}_n)d_n
    \label{eq:gn-te-cont}
\end{equation}
where
\begin{multline}
    \mathcal{C}_n x = \T{j}^{-1}\mu_\T{r}^{-1}\sqrt{\varepsilon_\T{r}}^{-1}\\\times\left[\sqrt{\varepsilon_\T{r}}\Hh_n(ka)\Jh_n'(k_\T{d}a) -\sqrt{\mu_\T{r}}\Hhp_n(ka)\Jh_n(k_\T{d}a)\right]x
    \label{eq:cnx}
\end{multline}
and
\begin{multline}
    \mathcal{D}_n x = \omega\sqrt{\varepsilon_\T{r}\mu_\T{r}}^{-1}\Hh_n(ka)\\\times\int_{-\infty}^\infty \omega^{\circ-1}\eta\hat{\sigma}(\omega-\omega^\circ)x(\omega^\circ)\Jh_n(k^\circ_\T{d}a) \T{d}\omega^\circ.
\end{multline}

\input{equations/b-and-c-matrix-eqs}

\section{Fourier series expansion of surface parameters}
\label{sec:fourier}

The preceding formulation involves integral operators whose kernels depend on the frequency domain representation of the surface conductivity $\hat{\sigma}(\omega)$.  In order to systematically solve these equations, we consider the special case in which the time-variation of the surface conductivity is periodic.  The aim of this simplification is to recast \eqref{eq:gn-tm-cont} and \eqref{eq:gn-te-cont} as matrix equations, similar to the construction of conversion matrices in time-varying circuit analysis \cite{maas2003nonlinear}.

Let the surface conductivity be expressible in terms of a Fourier series with fundamental frequency $\omega_\sigma$, i.e.,
\begin{equation}
    \hat{\sigma}(\omega) = \sum_{q=-K}^{K}\hat{\sigma}^q\delta(\omega-q\omega_\sigma).
\end{equation}
Substitution into the operator $\mathcal{B}_n$ gives, via the sifting property of the Dirac delta,
\begin{multline}
    \mathcal{B}_n x = \omega\eta\sqrt{\varepsilon_\T{r}\mu_\T{r}}^{-1}\Hhp_n(\omega a/c)\\\times\sum_{q=-K}^{K}\hat{\sigma}^q (\omega-q\omega_\sigma)^{-1}\Jh_n'((\omega-q\omega_\sigma)a/c_\T{d}) x(\omega-q\omega_\sigma),
\end{multline}
where $c$ and $c_\T{d}$ are the speed of light in vacuum and the dielectric core, respectively.  Further, assume that the excitation has a center frequency $\omega_0$ and a baseband representation that is periodic in the fundamental frequency $\omega_\sigma$, i.e.,
\begin{equation}
    g_n(\omega) = \sum_{p = -K}^K g_n^p\delta(\omega-\omega_0-p\omega_\sigma).
\end{equation}
Using these assumptions, \eqref{eq:gn-tm-cont} evaluated at frequency $\omega_p = \omega_0 + p\omega_\sigma$ may be written,
\begin{equation}
    g_n^p = A_n^{p}c_n^p +\sum_{q=-K}^K B_n^{pq}c_n^{p-q},
\end{equation}
with $c_n^p = c_n(\omega_p)$,
\begin{multline}
    A_n^{p} = \T{j}^{-1}\mu_\T{r}^{-1}\sqrt{\varepsilon_\T{r}}^{-1}\biggl[\sqrt{\mu_\T{r}}\Hh_n(k^p a)\Jh_n'(k^p_\T{d}a) \\ -\sqrt{\varepsilon_\T{r}}\Hhp_n(k^p a)\Jh_n(k^p_\T{d}a)\biggr],
    \label{eq:anp}
\end{multline}
and
\begin{multline}
    B_n^{pq} =  \omega_p\eta\sqrt{\varepsilon_\T{r}\mu_\T{r}}^{-1} \Hhp_n(k^p a)\hat{\sigma}^q\Jh_n'(k^{p-q}_\T{d}a)\omega_{p-q}^{-1}
    \label{eq:bnpq}
\end{multline}
where $k^p$ and $k^p_\T{d}$ are the wavenumbers at frequency $\omega_p$ in the exterior and core materials, respectively.  A set of $2K+1$ equations of this form may be collected into the system 
\begin{equation}
    \M{g}_n = \left(\M{A}_n + \M{B}_n\right)\M{c}_n
    \label{eq:g-vector-tm}
\end{equation}
where
\begin{equation}
    \M{A}_n = \begin{bmatrix}
    A_n^{-K} & 0 & 0 & 0 \\
    0 & A_n^{-K+1} & 0 & 0\\
    0 & 0 & \ddots & 0\\
    0 & 0 & 0 & A_n^K
    \end{bmatrix}
\end{equation}
and the matrix $\M{B}_n$ is given in factored form in \eqref{eq:bmatrix-long}.  This factorization is most easily constructed from \eqref{eq:bnpq} by substituting $\beta = p-q$ and inspecting the elements $B_n^{p\beta}$.

\setcounter{equation}{42}
Similarly for the TE coefficients we may write
\begin{equation}
    \M{g}_n = \left(\M{C}_n + \M{D}_n\right)\M{d}_n
    \label{eq:g-vector-te}
\end{equation}
where
\begin{equation}
    \M{C}_n = \begin{bmatrix}
    C_n^{-K} & 0 & 0 & 0 \\
    0 & C_n^{-K+1} & 0 & 0\\
    0 & 0 & \ddots & 0\\
    0 & 0 & 0 & C_n^K
    \end{bmatrix}
\end{equation}
\setcounter{equation}{44}
with elements analogous to \eqref{eq:anp} based on \eqref{eq:cnx} and the matrix $\M{D}_n$ as given in \eqref{eq:dmatrix-long}\stepcounter{equation}.  The form of the inner matrix
\begin{equation}
    \bs{\sigma} = \begin{bmatrix}
    \hat{\sigma}^{0} &  \hat{\sigma}^{-1} & \hdots & \hat{\sigma}^{-2K} \\
    \hat{\sigma}^{1} &  \hat{\sigma}^0 & \hdots & \hat{\sigma}^{-2K+1} \\
    \vdots & \vdots & \vdots & \vdots\\
    \hat{\sigma}^{2K} &  \hat{\sigma}^{2K-1} & \hdots & \hat{\sigma}^0 \\
    \end{bmatrix}
\end{equation}
closely resembles that of the conversion matrix describing multi-harmonic coupling in time-varying circuit analysis \cite{maas2003nonlinear} or method of moments problems involving time-varying loads and materials \cite{bass2021conversion}.  Interpretation of this quantity as the conversion matrix representation of a time-varying admittance is discussed further in Sec.~\ref{sec:tl}.

Together \eqref{eq:g-vector-tm} and \eqref{eq:g-vector-te} allow for the calculation of the internal Fourier components $\M{c}_n$ and $\M{d}_n$, from which the external terms $\M{a}_n$ and $\M{b}_n$ are obtained via \eqref{eq:an-cn} and \eqref{eq:bn-dn}, respectively.  Adopting the alternative modal normalization of Bohren and Huffman~\cite{bohren2008absorption}, we define new sets of external coefficients
\begin{equation}
    \tilde{\M{a}}_n = -\M{a}_n\frac{n(n+1)}{\T{j}^{-n}(2n+1)},\quad \tilde{\M{b}}_n = -\M{b}_n\frac{n(n+1)}{\T{j}^{-n}(2n+1)}.
\end{equation}
Assuming a monochromatic incident wave of power density $S_0$ at frequency $\omega_0$, this renormalization leads to a familiar form for the normalized net scattered power (scattering efficiency) at each frequency,
\begin{equation}
    Q_\T{sc}^p = \frac{P_\T{sc}^p}{S_0 \pi a^2} = \frac{2c^2}{\omega_p^2a^2}\sum_{n=1}^\infty (2n+1)\left[|\tilde{a}_n^p|^2 + |\tilde{b}_n^p|^2\right].
\end{equation}
The normalized extinction power (extinction efficiency) takes on a similar form
\begin{equation}
    Q_\T{ext}^0 = \frac{P_\T{ext}^0}{S_0 \pi a^2} = \frac{2c^2}{\omega_0^2a^2}\sum_{n=1}^\infty (2n+1)\T{Re}\,\left\{\tilde{a}_n^0 + \tilde{b}_n^0\right\}.
\end{equation}
By assuming monochromatic incidence, the extinction efficiency is only defined for the $p=0$ harmonic.  Because the time-varying scattering processes considered here are linear\footnote{See \cite[\S 3.4.1]{maas2003nonlinear} for extended discussion of conversion matrix linearity in the context of circuit problems.}, extended definitions of polychromatic scattering and extinction efficiencies may be constructed similarly to those used to describe polarization conversion processes in scattering cross section analyses.

\section{Surface resistance and conductivity}

The time-variation of the surface conductivity $\sigma(t)$ may be alternatively described using surface resistance $r(t) = \sigma(t)^{-1}$.  Note that simple time-variations in one parameter may be problematic or ill-defined in terms of the other.  For example, the simple time-varying surface resistance
\begin{equation}
    r(t) = r_0\left(1+\gamma\cos \omega_\sigma t\right),\quad \gamma\leq 1
    \label{eq:r-single-harm}
\end{equation}
has a well-behaved Fourier spectrum, however the associated surface conductivity tends toward a train of Dirac delta functions as $\gamma \rightarrow 1$ and $r_0\rightarrow \infty$.  This, in turn, necessitates the use of extreme numbers of frequencies in the solution of scattered fields which may lead to non-convergent results.  Throughout this paper we consider examples using conductance and resistance of the above form to examine this behavior.

\newcommand{\upia}{\V{u}_\alpha^{(1)}}
\newcommand{\upiab}{\V{u}_{\bar{\alpha}}^{(1)}}
\newcommand{\uppa}{\V{u}_\alpha^{(\rbkind)}}
\newcommand{\uppab}{\V{u}_{\bar{\alpha}}^{(\rbkind)}}
\newcommand{\upoa}{\V{u}_\alpha^{(4)}}
\newcommand{\upoab}{\V{u}_{\bar{\alpha}}^{(4)}}
\newcommand{\Ra}{\T{R}_\alpha}
\newcommand{\Rb}{\T{R}_{\bar{\alpha}}}

\section{Transition matrix formulation}
\label{sec:tmat}
\input{equations/y-matrix-equation}

In preceding sections we explicitly considered incident fields of the form of plane waves.  The formulation may be extended to more general illuminations through the use of transition (T-) matrix formalism \cite{waterman1965matrix,chew1990inhomogeneous}.  Consider an incident field of the form
\begin{equation}\label{eq:einc}
    \V{E}^\T{i}(\omega,\V{r}) = \sum_\alpha \upia(k\V{r}) g_\alpha
\end{equation}
where $\upia$ are vector spherical harmonics~\cite{ScatteringofEMWavesbyObstacles} associated with Riccati-Bessel functions of the first kind, see App.~\ref{sec:special-functions}.  By the addition theorem for vector spherical waves, this expansion is capable of representing any incident field produced by sources at locations $\V{r}'$ outside of the region of interest, i.e., $r<r'$~\cite[Eq. 7.5.14]{jin2011theory}. In a similar manner, the scattered and internal fields may be expanded as
\begin{equation}\label{eq:escat}
    \V{E}^\T{s}(\omega,\V{r}) = \sum_\alpha \upoa(k\V{r}) f_\alpha
\end{equation}
\begin{equation}\label{eq:ediel}
    \V{E}^\T{d}(\omega,\V{r}) = \sum_\alpha \upia(k_\T{d}\V{r}) h_\alpha
\end{equation}
where $\upoa$ are vector spherical harmonics associated with Riccati-Hankel functions of the second kind.  The property \cite[Eq. 7.2.4]{chew1990inhomogeneous}
\begin{equation}
    \nabla\times\uppa = k\uppab
\end{equation}
relates two classes of spherical vector waves with dual superindices $\alpha$ and $\bar{\alpha}$ \cite{losenicky2020method}.  By virtue of this property, the magnetic fields associated with incident, scattered, and internal electric fields are given by 
\begin{equation}\label{eq:hinc}
    \V{H}^\T{i}(\omega,\V{r}) = \T{j}\eta^{-1}\sum_\alpha \upiab(k\V{r}) g_\alpha
\end{equation}
\begin{equation}\label{eq:hscat}
    \V{H}^\T{s}(\omega,\V{r}) = \T{j}\eta^{-1}\sum_\alpha \upoab(k\V{r}) f_\alpha
\end{equation}
\begin{equation}\label{eq:hdiel}
    \V{H}^\T{d}(\omega,\V{r}) = \T{j}\eta_\T{d}^{-1}\sum_\alpha \upiab(k_\T{d}\V{r}) h_\alpha.
\end{equation}

Tangential electric field continuity in \eqref{eq:e-theta-cont} and \eqref{eq:e-phi-cont} at the sphere's boundary leads to
\begin{equation}
    g_\alpha\Ra^{(1)}(ka) + f_\alpha \Ra^{(4)}(ka) = h_\alpha \Ra^{(1)}(k_\T{d}a),
    \label{eq:tmat-e-cont}
\end{equation}
where $\Ra^{(p)}$ describe the radial dependence of spherical waves $\uppa$, see App.~\ref{sec:special-functions}.  Similarly, the magnetic field boundary conditions in \eqref{eq:h-theta-cont} and \eqref{eq:h-phi-cont} lead to 
\begin{multline}
    g_\alpha \Rb^{(1)}(ka) + f_\alpha\Rb^{(4)}(ka) = \frac{\eta}{\eta_\T{d}}h_\alpha \Rb^{(1)}(k_\T{d}a) \\- \T{j}\eta\gamma_\alpha\hat{\sigma}\star\left[h_\alpha\Ra^{(1)}(k_\T{d}a)\right]
    \label{eq:tmat-h-cont}
\end{multline}
where the identity
\begin{equation}\label{eq:signconstdefn}
    \u{r}\times\left(\u{r}\times\V{u}^{(\rbkind)}_{\bar\alpha}(\u{r})\right) = \u{r}\times\gamma_\alpha\V{u}^{(\rbkind)}_{\alpha}(\u{r}),\quad \gamma_\alpha = -1^{\tau}
\end{equation}
has been employed to relate tangential field components of harmonics with dual superindices. Note that the index $\tau$ is contained within the superindex $\alpha$, see Appendix A. 
Eliminating the external coefficients $f_\alpha$ gives 
\begin{equation}
    g_\alpha = (\Xop_\alpha+\Yop_\alpha)h_\alpha
    \label{eq:tmat-op-eq}
\end{equation}
where the operators $\Xop_\alpha$ and $\Yop_\alpha$ have the forms
\begin{multline}
    \Xop_\alpha x = \\\mathcal{W}_\alpha^{-1}\left[\frac{\eta}{\eta_\T{d}}\Ra^{(4)}(ka)\Rb^{(1)}(k_\T{d}a) - \Rb^{(4)}(ka)\Ra^{(1)}(k_\T{d}a)\right]x
    \label{eq:w-alpha}
\end{multline}
\begin{multline}
    \Yop_\alpha x = \\
    -\T{j}\eta \mathcal{W}_\alpha^{-1}\gamma_\alpha\Ra^{(4)}\int_{-\infty}^\infty \hat{\sigma}(\omega-\omega^\circ)x(\omega^\circ)\Ra^{(1)}(k^\circ_\T{d}a) \T{d}\omega^\circ
\end{multline}
with
\begin{equation}
    \mathcal{W}_\alpha x = \left[\Rb^{(1)}(ka)\Ra^{(4)}(ka) - \Ra^{(1)}(ka)\Rb^{(4)}(ka)\right]x.
    \label{eq:wdef}
\end{equation}
Adopting the Fourier series representations used in Sec.~\ref{sec:fourier}, we may convert the continuous operator equation in \eqref{eq:tmat-op-eq} into a matrix equation of the form
\begin{equation}
    \M{g}_\alpha = \left(\Xmat_\alpha + \Ymat_\alpha\right)\M{h}_\alpha
    \label{eq:tmat-mat-eq}
\end{equation}
where the matrix $\Xmat_\alpha$ is diagonal with elements
\begin{multline}
    \Xopchar_\alpha^{p} = W^{p,-1}_\alpha\\
    \times\left[\frac{\eta}{\eta_\T{d}}\Ra^{(4)}(k^pa)\Rb^{(1)}(k^p_\T{d}a) - \Rb^{(4)}(k^pa)\Ra^{(1)}(k^p_\T{d}a)\right]
\end{multline}
and the matrix $\Ymat_\alpha$ is given in factored form in \eqref{eq:ymatrix-long}\stepcounter{equation}.  Here the term $W^{p}_\alpha$ corresponds to the coefficient in \eqref{eq:wdef} evaluated at frequency $\omega_p$.  
After some manipulations, this leads to a transition matrix system
\begin{equation}
    \M{f}_\alpha = \M{T}_\alpha\M{g}_\alpha
    \label{eq:tmat-def}
\end{equation}
where
\begin{equation}
    \M{T}_\alpha = \M{R}_{\alpha,4}^{-1}\left[\M{R}_{\alpha,1\T{d}}\left(\Xmat_\alpha+\Ymat_\alpha\right)^{-1}-\M{R}_{\alpha,1}\right]
    \label{eq:t-mat-final}
\end{equation}
with
\begin{equation}
    \M{R}_{\alpha,\rbkind} = \begin{bmatrix}
    \Ra^{(\rbkind)}(k^{-K}a) & 0 & 0 \\
    0  & \ddots & 0\\
    0  & 0 & \Ra^{(\rbkind)}(k^{K}a)
    \end{bmatrix}
    \label{eq:rp-mat}
\end{equation}
and    
\begin{equation}
    \M{R}_{\alpha,\rbkind\T{d}} = \begin{bmatrix}
    \Ra^{(\rbkind)}(k_\T{d}^{-K}a) & 0 & 0 \\
    0  & \ddots & 0\\
    0  & 0 & \Ra^{(\rbkind)}(k_\T{d}^{K}a)
    \end{bmatrix}.
    \label{eq:rpd-mat}
\end{equation}
The transition matrix in \eqref{eq:tmat-def} fully describes the scattering of multi-harmonic incident fields originating from sources exterior to the sphere under consideration.  Hence it is suitable, through proper application of the addition theorem, to adaptation toward the scattering analysis of collections of non-overlapping spherical structures.

\section{Spherical waveguide perspective}
\label{sec:tl}
Here we consider an alternative transmission line perspective of the scattering problem presented in Secs.~\ref{sec:pw}, \ref{sec:fourier}, and~\ref{sec:tmat}. In general, we may consider the free space around the sphere as a spherical waveguide, in which each mode has an associated characteristic  admittance \cite{harrington2001time}. 
The wave admittance $Y_\alpha^{(\rbkind)}$ of a spherical wave defined by an electric field of the form $\V{u}_\alpha^{(\rbkind)}$ relates the magnitudes of the tangential electric and magnetic fields via
\begin{equation}
    \V{H}^{\T{tan}}_\alpha = Y^{(\rbkind)}_\alpha \u{r}\times\V{E}^{\T{tan}}_\alpha.
\end{equation}
Using \eqref{eq:einc} and \eqref{eq:hinc}, \eqref{eq:escat} and \eqref{eq:hscat}, or \eqref{eq:ediel} and \eqref{eq:hdiel} for the incident, scattered, or internal fields respectively, we define a diagonal multi-frequency wave admittance matrix
\begin{equation}
    \M{Y}_{\alpha,\rbkind m} = \T{j}\gamma_\alpha\eta_m^{-1}\M{R}_{\bar{\alpha}, \rbkind m} (\M{R}_{\alpha, \rbkind m})^{-1},
    \label{eq:admittance-mat}
\end{equation}
where, following the notation of \eqref{eq:rp-mat} and \eqref{eq:rpd-mat}, the subscript $\rbkind m$
~may be either $\rbkind$ for waves of the $\rbkind$th kind in free space or $\rbkind\T{d}$ for waves in the dielectric core. Similarly, the impedance $\eta_m$ is either that of free space or the dielectric core. The sign constant $\gamma_\alpha$ is $\pm 1$ as given in \eqref{eq:signconstdefn}, which produces an appropriate sign for TE and TM modes.

The electric field continuity equation of \eqref{eq:tmat-e-cont} may be written in terms of incident, reflected, and transmitted ``voltages''
~at each frequency as 
\begin{equation}
    v^\T{i}_\alpha + v^\T{r}_\alpha = v^\T{t}_\alpha,
    \label{eq:tline-e-cont}
\end{equation}
where the incident voltages are
\begin{equation}
    v^\T{i}_\alpha = g_\alpha\Ra^{(1)}(ka),
\end{equation}
and similar forms are used for the reflected and transmitted terms.  Collecting equations of this kind at all frequencies we obtain the system of equations
\begin{equation}
    \M{v}_\alpha^\T{i} + \M{v}_\alpha^\T{r} = \M{v}^\T{t}_\alpha.
\end{equation}

Using the same voltage definitions along with admittance matrices of the form of \eqref{eq:admittance-mat}, the magnetic field  continuity equation of \eqref{eq:tmat-h-cont} may be written as
\begin{equation}\label{eq:TLcurrent2}
    \M{Y}_{\alpha,1} \M{v}^\T{i}_\alpha   + \M{Y}_{\alpha,4} \M{v}^\T{r}_\alpha  =  \M{Y}_{\alpha,1\T{d}}\M{v}^\T{t}_\alpha + \V{\sigma}\M{v^\T{t}}_\alpha,
\end{equation}
where the convolution term in \eqref{eq:tmat-h-cont} is represented by cross-frequency (off-diagonal) terms in the conversion matrix $\V{\sigma}$.

\begin{figure}
    \centering
    \input{figures/fig-02-tline}
    \caption{Schematic of transmission line scattering model.  The left and right transmission lines represent the exterior and interior regions, respectively, while the shunt conductance represents the surface conductivity.}
    \label{fig:TLschematic}
\end{figure}
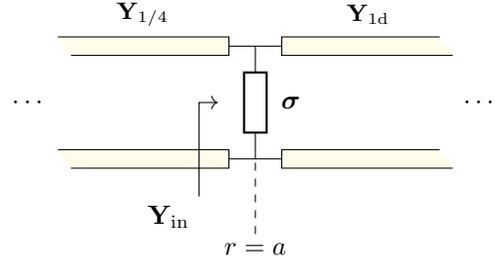

Interpreting \eqref{eq:tmat-e-cont} and \eqref{eq:TLcurrent2} as the voltage and current boundary conditions at a transmission line junction, it is clear that the surface of the sphere appears as a time-varying conductance in parallel with the input admittance ``looking into'' the core.  This configuration is depicted in Fig.~\ref{fig:TLschematic}, where a compressed notation is used to depict wave impedances in each region.  Because of the orthogonality of vector spherical harmonics, no mode conversion occurs at the surface of the sphere, but the time-varying boundary condition does result in frequency conversion. Thus, while each harmonic frequency satisfies \eqref{eq:tline-e-cont} independently, \eqref{eq:TLcurrent2} relates multiple harmonic frequencies simultaneously and cannot be separated into an independent equation for each frequency.

We now set out to derive the scattered modal voltage $\M{v}^\T{r}_\alpha$ using classical transmission line techniques.  Let 
\begin{equation}\label{eq:TLgammadefn}
    \M{v}^\T{r}_\alpha = \M{\Gamma}_\alpha \M{v}^\T{i}_{\alpha},
\end{equation}
where $\M{\Gamma}_\alpha$ is a multi-frequency reflection coefficient matrix that relates the incident and scattered wave amplitudes in the $\alpha$ mode at all frequencies.
By combining \eqref{eq:tline-e-cont}, \eqref{eq:TLcurrent2}, and \eqref{eq:TLgammadefn} and eliminating $\M{v}^\T{i}_\alpha$, we obtain
\begin{equation}
   \M{Y}_{\alpha, \T{in}} \left(\V{1}+\M{\Gamma}_\alpha\right)
    = (\M{Y}_{\alpha,1} + \M{Y}_{\alpha,4}\M{\Gamma}_\alpha ),
\end{equation}
where the input admittance conversion matrix looking into the combined core-shell structure at each frequency is
\begin{equation}
    \M{Y}_{\alpha, \T{in}} = \M{Y}_{\alpha,1\T{d}} +\V{\sigma}.
\end{equation}
Collecting terms and solving for the reflection coefficient matrix $\M{\Gamma}_\alpha$ we obtain
\begin{equation}
    \M{\Gamma}_\alpha = (\M{Y}_{\alpha,4}-\M{Y}_{\alpha, \T{in}})^{-1}(\M{Y}_{\alpha, \T{in}}- \M{Y}_{\alpha,1}).
    \label{eq:gamma-mat-final}
\end{equation}

The expression for the multi-frequency reflection coefficient matrix in \eqref{eq:gamma-mat-final} has the same form as that arising from a time-invariant transmission line interface, with the major distinction that \eqref{eq:gamma-mat-final} represents a multi-harmonic system with conversion between frequencies. Again we note that the incident and outward-going wave admittances are diagonal matrices, while the input admittance matrix has off diagonal elements which account for frequency conversion due to the time-varying conductivity.  Additionally, since the incident wave is represented by a standing (Bessel) wave whose nominal propagation direction is chosen outward, the sign of the magnetic field is the same for all wave components so the ``numerator'' and ``denominator'' both appear as subtractions.  Lengthy manipulations, outlined in App.~\ref{sec:app-tl-and-tmat}, demonstrate an equivalence between the result in \eqref{eq:gamma-mat-final} and that obtained via transition matrix formulation in \eqref{eq:t-mat-final}.  The intuitive construction of the equivalent circuit in Fig.~\ref{fig:TLschematic} suggests that the study of more complex systems, e.g., layered media or shells, may be facilitated through the use of similar transmission line methods.

\section{Convergence and comparison to other methods}
\label{sec:ex-1}

Here we consider two sets of examples aimed at examining the accuracy and convergence of the methods presented in this work.  For brevity, we restrict our analyses to those based on the special case of plane wave incidence, outlined in Sec.~\ref{sec:pw}.

\subsection{Comparison to numerical methods, air-core shell}

When the dielectric core material is assigned to be the same as the exterior medium, the plane wave scattering problem in Sec.~\ref{sec:pw} is readily evaluated using a hybridized conversion matrix method of moments (CMMoM) technique~\cite{bass2021conversion} developed for the study of conducting structures loaded with time-varying lumped elements or materials. Here we compare these purely numerical results with those obtained using the analytical formulation presented in this work. 

As an example problem, we consider an air-core conducting shell with conductivity of the form
\begin{equation}
\sigma(t) = \sigma_0(1+\gamma\cos\omega_\sigma t)
\label{eq:sigma-def}
\end{equation}
where $\sigma_0 = 1~\Omega^{-1}$, $\gamma = 0.5$, and $\omega_\sigma = 0.11\omega_0$.   In contrast to the example resistivity given in \eqref{eq:r-single-harm}, this conductivity results in surface resistivity of the form
\begin{equation}
    r(t)=(\sigma_0(1+\gamma\cos\omega_\sigma t))^{-1}.
\end{equation}

The incident field is a monochromatic plane wave~ whose frequency $\omega_0$ is  swept to cover a broad range of electrical sizes $ka$.  Scattering and extinction efficiencies are calculated via the method presented in Sec.~\ref{sec:pw}, and via CMMoM where the sphere is represented by 900 RWG basis functions and the time-varying conductivity is modeled as a distributed time-varying surface resistance.  In the CMMoM implementation, all impedance and Gram matrices are generated using AToM~\cite{atom}.  CMMoM scattering and extinction efficiencies are calculated via
\begin{equation}
    Q_\T{sc}^p = \frac{P_\T{sc}^p}{S_0 \pi a^2} = \frac{\M{I}_p^\T{H}\M{R}_p\M{I}_p}{2S_0\pi a^2}
\end{equation}
and
\begin{equation}
   Q_\T{ext}^p =\frac{P_\T{ext}^p}{S_0 \pi a^2} = \frac{\T{Re}\,\{\M{I}_p^\T{H}\M{V}_p\}}{2S_0\pi a^2}
\end{equation}
where $\M{I}_p$, $\M{V}_p$, and $\M{R}_p$ are the induced current, excitation field, and radiation operator at the $p^{th}$ harmonic represented in the MoM basis.

\begin{figure}
    \centering
    \includegraphics[width=3.25in]{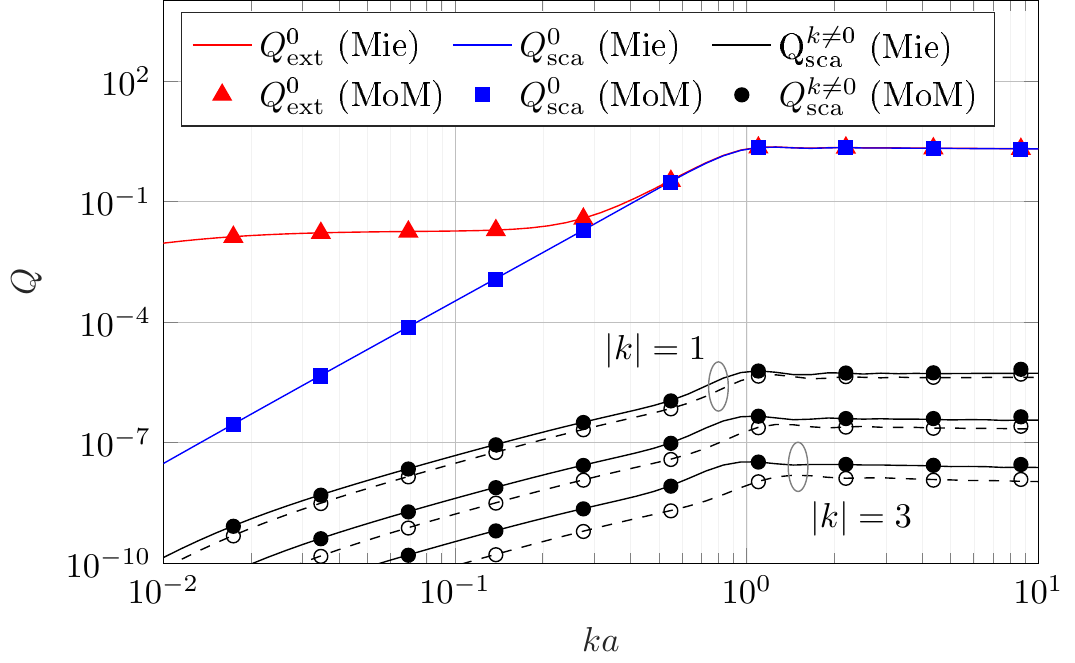}
    \caption{Comparison of scattering and extinction efficiencies calculated by Mie (this paper) and CMMoM formulations with $\varepsilon_\T{r} = \mu_\T{r} = 1$ and $\sigma(t)$ of the form of \eqref{eq:sigma-def} where $\sigma_0 = 1~\Omega^{-1}$ and $\gamma = 0.5$.  At all frequencies, $\omega_\sigma = 0.11\omega$.  Red (blue) markers and traces denote extinction (scattering) at the excitation harmonic.  For harmonics with $k\neq 0$, solid black lines and markers indicate $k>0$, while dashed black lines and unfilled markers denote $k<0$.}
    \label{fig:mie-mom-comparison}
\end{figure}

Results obtained by the method presented in Sec.~\ref{sec:pw} (denoted ``Mie'') and CMMoM are shown in Fig.~\ref{fig:mie-mom-comparison}, where qualitative agreement is observed.  At all frequencies, scattering in the excitation ($p=0$) harmonic far outweighs scattering in up- and down-converted ($p\neq0$) harmonics.  Both methods agree to within $4\%$ in all reported quantities for small electrical sizes $ka<5$. More significant discrepancies appear for larger electrical sizes where the number of CMMoM basis functions limits the ability to accurately represent intricate higher-order mode shapes, particularly in up-converted harmonics. 

The quasi-analytic method is much less computationally complex than a CMMoM solution.  This is due to the relatively low number of spherical harmonics required to represent fields and currents as compared to the high number of discrete basis functions used in CMMoM. On the other hand, the quasi-analytic solution is only applicable to canonical geometries. Parameters for general computational costs of both methods are listed in Tab.~1 along with example data from a single data point from Fig.~\ref{fig:mie-mom-comparison}.  As in Fig.~\ref{fig:mie-mom-comparison}, the number of spatial CMMoM basis functions is $N_\T{bf} = 900$ and $K=4$ is used to control the number of harmonic frequencies.  The number of spherical harmonics $N_\T{h} = 6$ used in the quasi-analytic solution was determined based on convergence of all harmonic efficiencies, and this number is observed to generally increase with increasing frequency. Though the quasi-analytic solution requires multiple linear systems to be constructed and solved, these matrices are many orders of magnitude smaller than the multi-harmonic CMMoM system matrix.  Note that construction of CMMoM impedance and Gram matrices using AToM~\cite{atom} dominates the overall CMMoM solution time.

\begin{table}[]
    \centering
    \caption{Computational cost and evaluation times for CMMoM and the quasi-analytic Mie approach.  Both methods are implemented in MATLAB on an Intel i7-8700 3.20 GHz CPU.  MoM impedance and Gram matrix data are calculated using AToM\cite{atom}.}
    
    \begin{tabular}{c|c|c|c|c}
        & \multicolumn{2}{c}{General} & \multicolumn{2}{|c}{$ka = 1$} \\\hline\hline
        & \textbf{CMMoM} & \textbf{Mie} & \textbf{CMMoM} & \textbf{Mie} \\
        \hline
        matrix dimension & $(2K+1)N_\T{bf}$&  $2K+1$ & 8100 & 9 \\
        \# matrices & $1$ & $2N_\T{h}$ & 1 & 12 \\\hline
        fill time (s)& - & - & $120$ & $0.002$ \\
        solution time (s) & - &- & $8.1$ & $<0.001$ \\\hline
        \textbf{total time (s)} & - &- & $\mathbf{130}$ & $\mathbf{0.005}$
    \end{tabular}
    ~\\
    \label{tab:my_label}
\end{table}

\subsection{Convergence characteristics}
As in circuit applications of conversion matrix methods, the number of temporal harmonics must be sufficiently high to ensure accurate field and efficiency calculations.  We expect the required number of temporal harmonics $2K+1$ to depend on many factors, and here we examine a few of these dependencies using two simple test cases.

\begin{figure}
    \centering
    \includegraphics[width=3.25in]{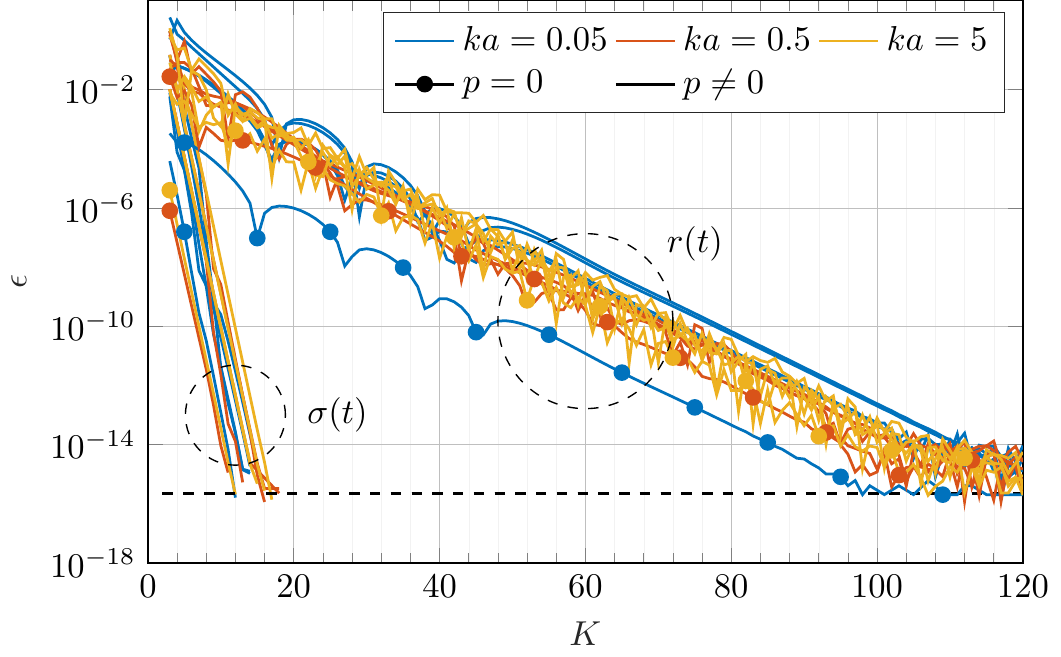}
    \caption{Convergence properties of scattering efficiencies with respect to increasing number of harmonics $K$ using a air-core sphere surrounded by a shell with 
    sinusoidally varying surface resistivity or conductivity as described by \eqref{eq:r-single-harm} or \eqref{eq:sigma-def}, respectively.  The electrical size $ka$ is given relative to the frequency of the incident plane wave. Traces with circular markers indicate the $p = 0$ fundamental frequency, while traces without markers represent harmonics with $p\neq 0$. } 
    \label{fig:conv-s0-easy}
\end{figure}

As an example of a system with relatively few harmonic components in the matrix $\V{\sigma}$, we consider an air-core sphere with surface conductivity of the form of \eqref{eq:sigma-def} where $\sigma_0 = 1~\Omega^{-1}$, $\omega_\sigma = 0.11\omega_0$, and $\gamma = 0.99$. The resulting matrix  $\V{\sigma}$ is tridiagonal.  The electrical size of the sphere is swept over the values $ka\in\{0.05,0.5,5\}$ and the multi-harmonic scattering efficiencies $Q_\T{sc}^p$ are recorded using increasing numbers of temporal harmonics $K$.  For all electrical sizes, a relative convergence error 
\begin{equation}
    \varepsilon(K) = \frac{|Q_\T{sc}^p(K)-Q_\T{sc}^p(K-1)|}{Q_\T{sc}^p(K-1)}  
\end{equation}
is calculated for each harmonic scattering efficiency.  Errors for all three electrical sizes are collected at each value of $K$ and plotted for the $p\in\{-2,...,2\}$ harmonics in Fig.~\ref{fig:conv-s0-easy}, labeled by a circle annotated with $\sigma(t)$.  These results indicate that, for all electrical sizes, this particular example converges rapidly to numerical precision for $K\approx 15$.

To model a system with much richer harmonic content, we additionally examine a sphere with surface resistance of the form of \eqref{eq:r-single-harm} with $\gamma = 0.99$, $r_0 = 500~\Omega$, and $\omega_\sigma = 0.11\omega$.  Convergence results from this analysis are also shown on Fig.~\ref{fig:conv-s0-easy}, labelled by a circle annotated with $r(t)$. In this case, the Fourier spectrum of the corresponding surface conductivity (obtained by inversion of the specified resistivity and truncated to $K$ harmonics) is no longer sparse and many more ($K\approx 100$) harmonics are required to reach numerical precision.  

Interestingly, the convergence properties of both problems show little dependence on the electrical size of the sphere. All harmonics of all cases for the $\sigma(t)$ model converge to numerical precision with approximately 15 harmonics, while the more harmonically complex $r(t)$ example requires about 100 harmonics. In general, the variation in error between harmonic frequencies for any given sphere size is larger than the variation between sphere sizes. For all sphere sizes, the $p=0$ harmonic has lower relative convergence error than most or all $p\neq 0$ harmonics of the same problem, though this effect is less pronounced at larger electrical sizes.

These examples illustrate the basic convergence properties of scattering from time-varying shells with simple resistance and conductance properties.  More importantly, however, they demonstrate the increase in complexity and computational cost when certain classes of surface conductivity Fourier spectra are used.  For analytical methods such as those presented in this paper, the increase in computational complexity is manageable, however this rapid scaling may quickly outpace computational resources in numerical methods, such as CMMoM, where multi-frequency, $N$-port impedance matrices with dimensions scaling linearly with both the number of harmonics and number of ports must be computed, stored, and inverted.

\section{Scattering behavior}
\label{sec:ex-2}
In Sec.~\ref{sec:ex-1}, examples were selected to demonstrate the accuracy or convergence of the method developed in this paper.  We now deviate from measures of accuracy and instead leverage the speed of these analytic methods to examine in detail the unique capabilities of scatterers with strong time-varying behavior.

\subsection{Study of surface and core characteristics}
\begin{figure}
    \centering
\includegraphics[width=3in]{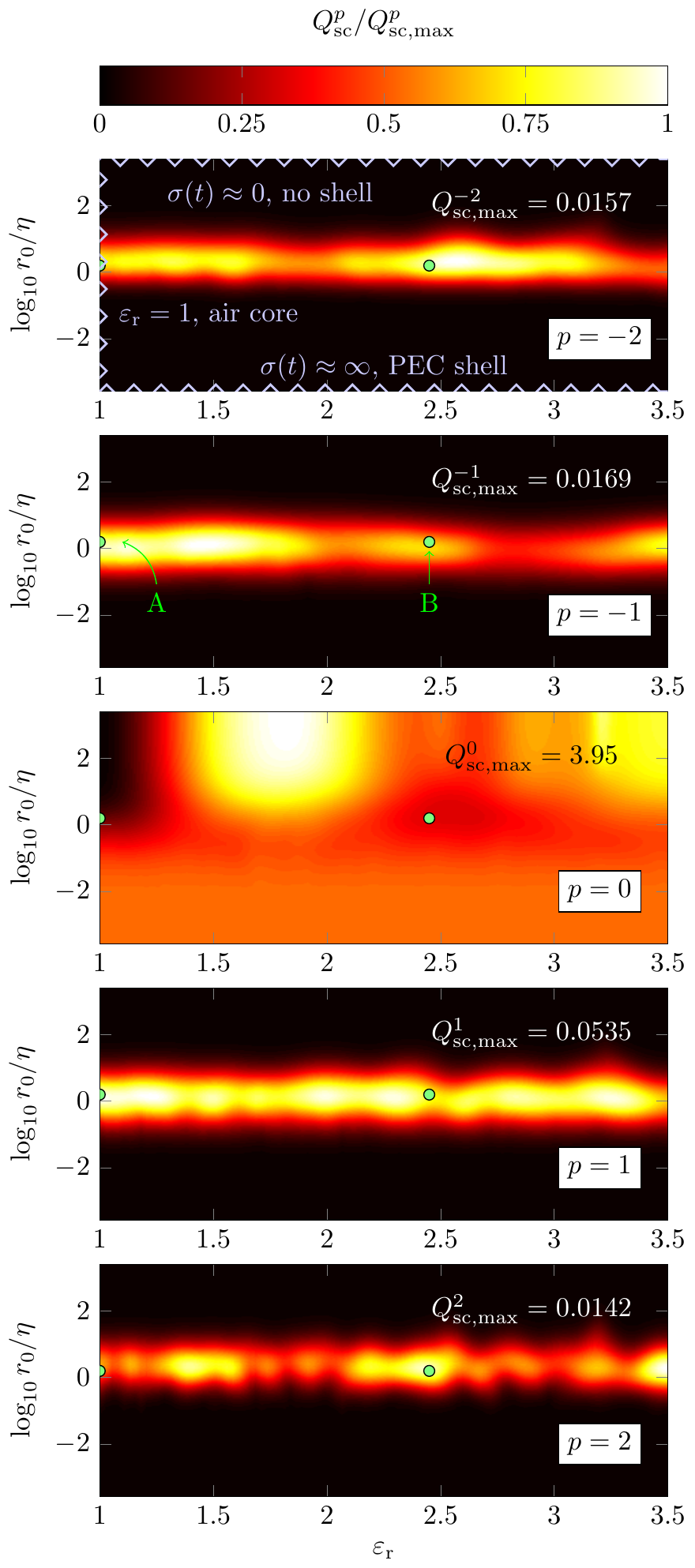}
    \caption{Scattering efficiencies $Q_\T{sc}^p$ for a dielectric core with relative permittivity $\varepsilon_\T{r}$ coated with a shell with time-varying resistance given by \eqref{eq:r-single-harm} with $\gamma = 0.9$, $\omega_\sigma = 1.5\omega$, and $ka = 2\pi$.}
    \label{fig:er-r0-sweep}
\end{figure}

Harmonic generation in the spherical system studied in this paper depends on many system parameters.  Here we maintain the simple surface resistance time dependence in \eqref{eq:r-single-harm} with $\gamma = 0.9$, $\omega_\sigma = 1.5\omega_0$, and $ka = 2\pi$ to examine the effect of material parameters, namely the core relative permittivity $\varepsilon_\T{r}$ and surface resistance $r_0$, on harmonic scattering. In Fig.~\ref{fig:er-r0-sweep}, we plot the scattering efficiencies $Q_\T{sc}^p$ for $p \in \{-2,...,2\}$ as functions of these two material parameters.  For reference, the limiting cases when the core is constructed of air ($\varepsilon_\T{r}=1$), the shell is PEC ($\sigma(t)\approx\infty$), and the shell is non-existent ($\sigma(t)\approx 0$) are indicated by zig-zag boundaries in the top panel of Fig.~\ref{fig:er-r0-sweep}.

From this study, we observe that harmonic generation is maximized near $r_0\approx\eta$ for all core permittivities, where $\eta$ is the characteristic impedance of free space.  Comparing scattering efficiencies in the air-core case as a function of resistivity $r_0$, we observe very similar trends as obtained in the study of an electrically small rectangular plate constructed of a similar time-varying conductor \cite{bass2021conversion}, with harmonic scattering efficiencies maximized near values of resistivity $r_0$ producing maximum net absorption. 

\begin{figure*}
\centering
\includegraphics[width=7in]{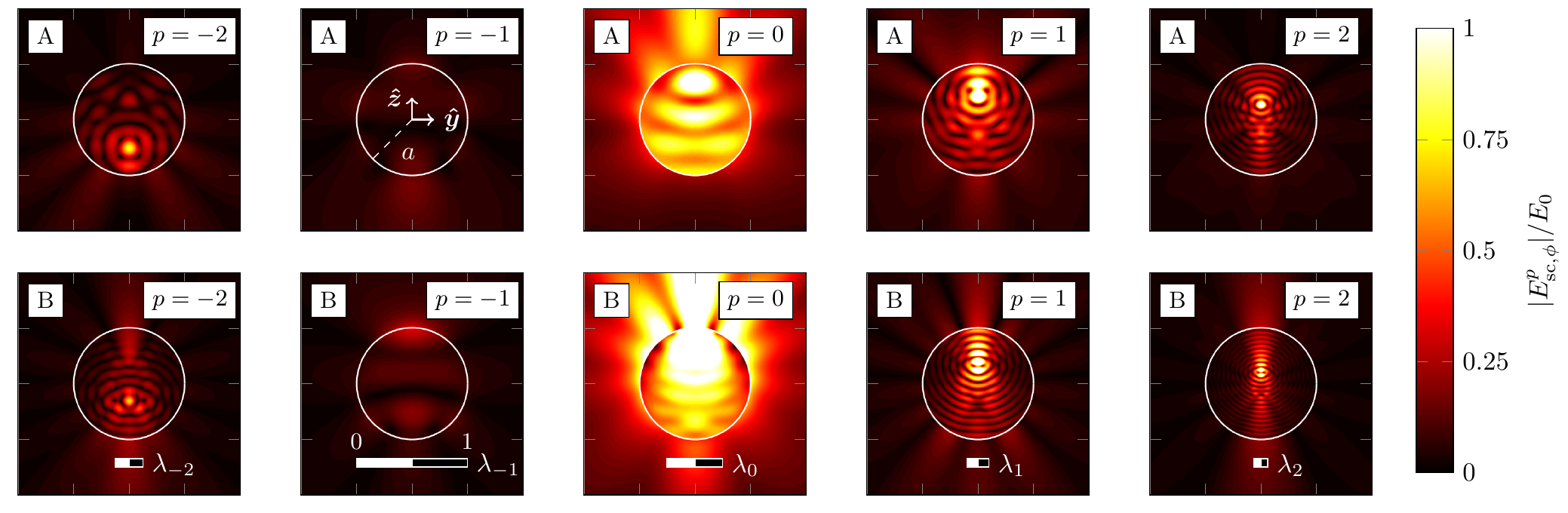}
\caption{Normalized scattered electric field $E^\T{s}_\phi$ magnitude produced by a sphere of radius $ka = 2\pi$ with core relative permittivity $\varepsilon_\T{r}$ and surface resistance defined by $r(t) = r_0(1+\gamma\cos\omega_\sigma  t)$ with $\omega_\sigma = 1.5\omega_0$ and $\gamma = 0.9$. Results for two pairs of parameters $(\varepsilon_\T{r},r_0)$ are shown corresponding to the labeled points ``A'' (top, $\varepsilon_\T{r} = 1, r_0/\eta = 2$) and ``B'' (bottom, $\varepsilon_\T{r} = 2.45, r_0/\eta = 2$) in Fig.~\ref{fig:er-r0-sweep}.  Scale bars in the bottom row indicate the free-space wavelength at each harmonic, valid for both rows of data.}
\label{fig:er-sweep-fields}
\vspace*{4pt}
\hrulefill
\vspace*{4pt}
\end{figure*}

Two particular sets of material parameters are labelled as points A and B within the middle panel of Fig.~\ref{fig:er-r0-sweep}.  The scattered electric field in each of the $p\in\{-2,...,2\}$ harmonics at these points are shown in Fig.~\ref{fig:er-sweep-fields}.  There, up- and down-conversion is clearly visible through the modification of standing wave spatial frequency, along with unique scattering behavior (e.g., directivity and overall intensity) within each harmonic.  Note that only the scattered field is shown.  In the $p=0$ excitation frequency this is not the total field and, as expected, there exists a field discontinuity at the boundary of the sphere.  For $p\neq0$ harmonics, the scattered field is the total field and no such discontinuity exists.

\subsection{Harmonic near-field generation}

\begin{figure}
    \centering
    \includegraphics[width=3.25in]{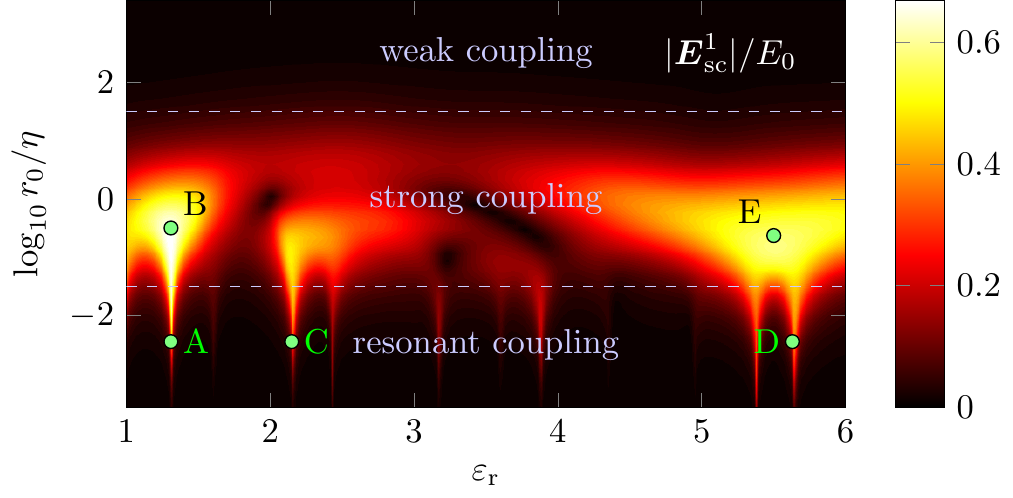}
    \caption{Electric near field magnitude at $(0,0,-a/2)$ in the $p = 1$ harmonic as a function of surface resistance magnitude $r_0$ and core relative permittivity $\varepsilon_\T{r}$ for a sphere of electrical size $ka = \pi/2$ at the excitation frequency, with $\gamma = 0.9$ and $\omega_\sigma = 1.5\omega$.}
    \label{fig:nf-sweep}
\end{figure}

\begin{figure}
    \centering
    \includegraphics[width=2.08in]{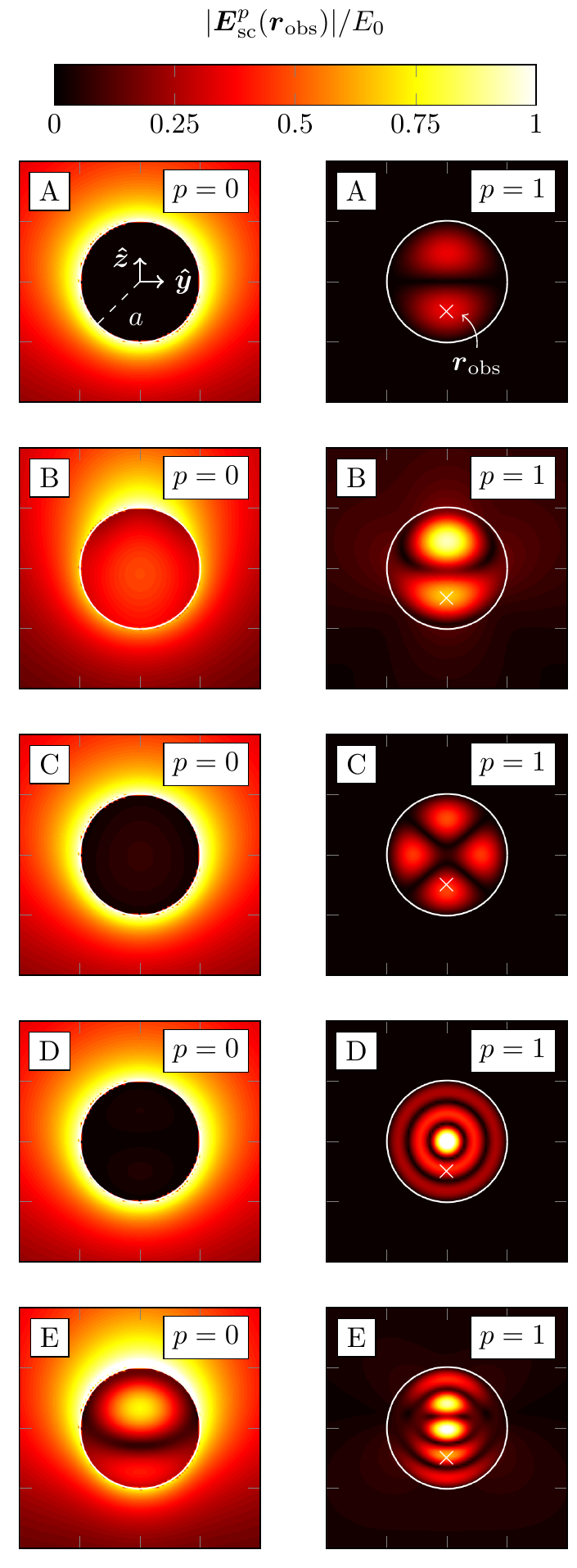}
    \caption{Scattered fields in the excitation ($p=0$) and first up-converted frequency ($p=1$) corresponding to the annotated positions in Fig.~\ref{fig:nf-sweep}.  All fields are normalized to the incident field magnitude.}
    \label{fig:nf-fields}
\end{figure}

Moving away from the study of scattering cross-sections, we now turn our attention toward harmonic near-field generation.  We consider a sphere of electrical size $ka=\pi/2$ coated with a resistive shell governed by \eqref{eq:r-single-harm} with $\gamma = 0.9$ and $\omega_\sigma = 1.5\omega_0$.  Defining an observation point $\V{r}_\T{obs} = -\V{\hat{z}}a/2$ within the core (see Fig.~\ref{fig:nf-fields} for schematic of observation location), we compute the normalized electric field magnitude $|E_\T{sc}^p|/E_0$ within each harmonic again as a function of core relative permittivity $\varepsilon_\T{r}$ and surface resistivity $r_0$.  This field magnitude for the first up-converted harmonic ($p=1$) is shown in Fig.~\ref{fig:nf-sweep}, where three distinct classes of behavior are apparent.  First, when the surface resistance $r_0$ is very low, the system behaves approximately as a PEC sphere with slight coupling between internal and external fields.  In this regime, we observe that the $p=1$ harmonic field magnitudes are maximized at the observation location over extremely narrow ranges of core permittivity corresponding to internal resonances of the sphere, hence the name \emph{resonant coupling} used to define this mode of operation.  As the surface resistance increases, the Q-factor of these resonances decreases, eventually leading to \emph{strong coupling} of multi-mode field distributions in the range $0.1\leq r_0/\eta \leq 10$.  Here the harmonics generated by the time-varying shell exist both inside the sphere and as outward propagating scattered fields.  As the resistance $r_0$ becomes very large, no currents are induced on the shell (making it effectively transparent) and harmonic fields tend toward zero.  Though not visible in Fig.~\ref{fig:nf-sweep}, in this \emph{weak coupling} regime, internal fields are maximized when the harmonic frequency of interest coincides approximately with the external resonances of the dielectric sphere in isolation.

In Fig.~\ref{fig:nf-fields}, field distributions at the excitation ($p=0$) and first harmonic ($p=1$) frequencies are shown for a selection of resistance and core permittivity values labeled as points A-E in Fig.~\ref{fig:nf-sweep}.  Selected for their high internal field magnitude and low values of resistance $r_0$, cases A, C, and D represent resonant coupling between the excitation field and the first harmonic internal fields.  In these cases, the first harmonic internal field distributions are effectively those corresponding to the internal resonances of a dielectric-filled PEC shell.  Because it is unlikely (and not the case in any of these examples) that the sphere has internal resonances at both the excitation and first harmonic frequencies, we observe that the internal fields at the excitation frequency are effectively zero.  In contrast, cases B and E exist in the strong coupling regime where the conductive shell is semi-transparent and significant internal and external fields exist at both excitation and harmonic frequencies.

\subsection{Trends in harmonic detuning}

\begin{figure}
    \centering
    \includegraphics[width=3.5in]{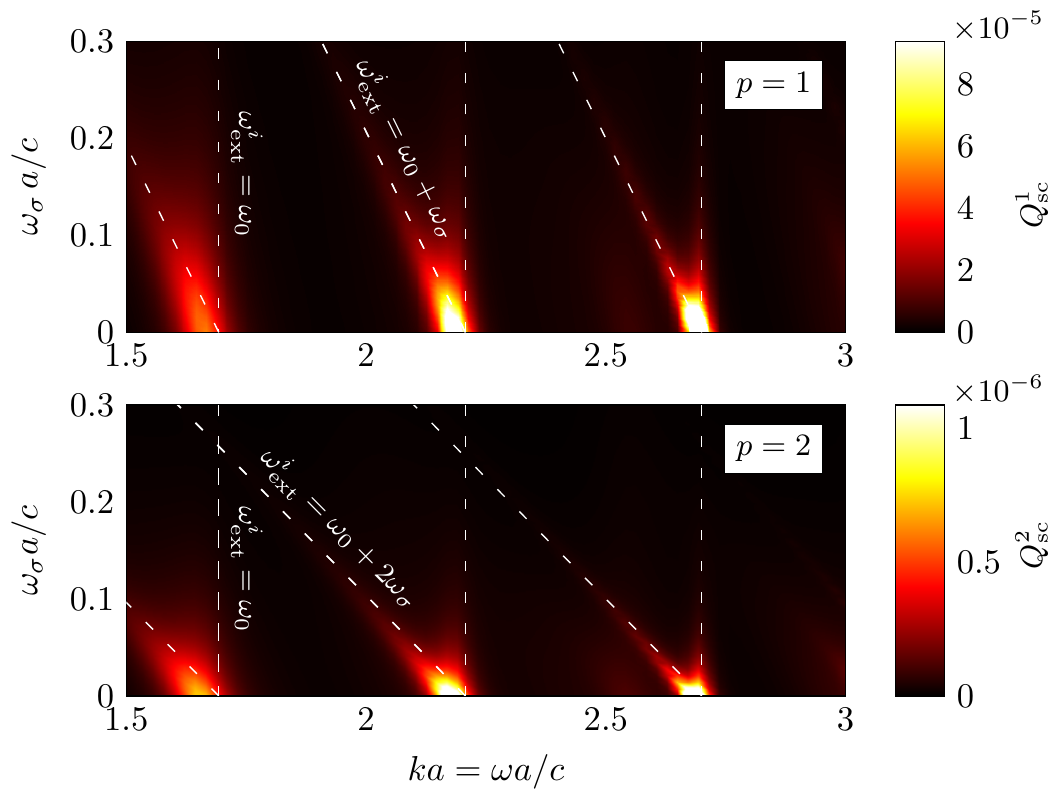}
    \caption{Scattering efficiency in the $p=1$ (top) and $p=2$ (bottom) harmonics from a sphere with relative core permittivity $\varepsilon_\T{r}=6.25$ and time-varying shell resistance given by \eqref{eq:r-single-harm} with $r_0/\eta = 10^2$ and $\gamma = 0.25$.}
    \label{fig:detuning-weak}
\end{figure}

Here we study the effective detuning of a sphere's natural internal and external resonances due to up-conversion processes enabled by a time-varying conductive shell.  Following the nomenclature of \cite{stefanou2021light}, we begin by defining external resonances of an uncoated dielectric sphere as the set of frequencies $\{\omega_\T{ext}^i\}$ producing maximal scattering efficiency.  In the weak coupling regime, where surface currents can be considered as a perturbation on an otherwise uncoated dielectric sphere, we anticipate that if energy is coupled into the harmonic $\omega_p$, the scattering efficiency $Q_\T{sc}^p$ will be maximized when that frequency corresponds to one of the external resonances of the dielectric core, i.e., when
\begin{equation}
    \omega_\T{ext}^i = \omega_p = \omega_0 + p\omega_\sigma
    \label{eq:external-resonance}
\end{equation}
is satisfied.  In Fig.~\ref{fig:detuning-weak}, we plot the scattering efficiency of the first ($p=1$) and second ($p=2$) upconverted harmonics as a function of excitation frequency $\omega$ and shell variation frequency $\omega_\sigma$ using a dielectric core with relative permittivity $\varepsilon_\T{r}=6.25$ and time-varying resistance following \eqref{eq:r-single-harm} with $r_0/\eta = 10^2$ and $\gamma = 0.25$.  The external resonance frequencies $\{\omega_\T{ext}^i\}$ are plotted as vertical dashed lines, while combinations of excitation and shell frequencies satisfying \eqref{eq:external-resonance} are drawn as diagonal dashed lines.  We observe that the maximum scattering efficiencies typically align with the condition in \eqref{eq:external-resonance}, though harmonic generation is enhanced particularly for cases with $\omega_\sigma \approx 0$ where both the harmonic frequency and the excitation frequencies are near an external resonance. 

\begin{figure}
    \centering
    \includegraphics[width=3.5in]{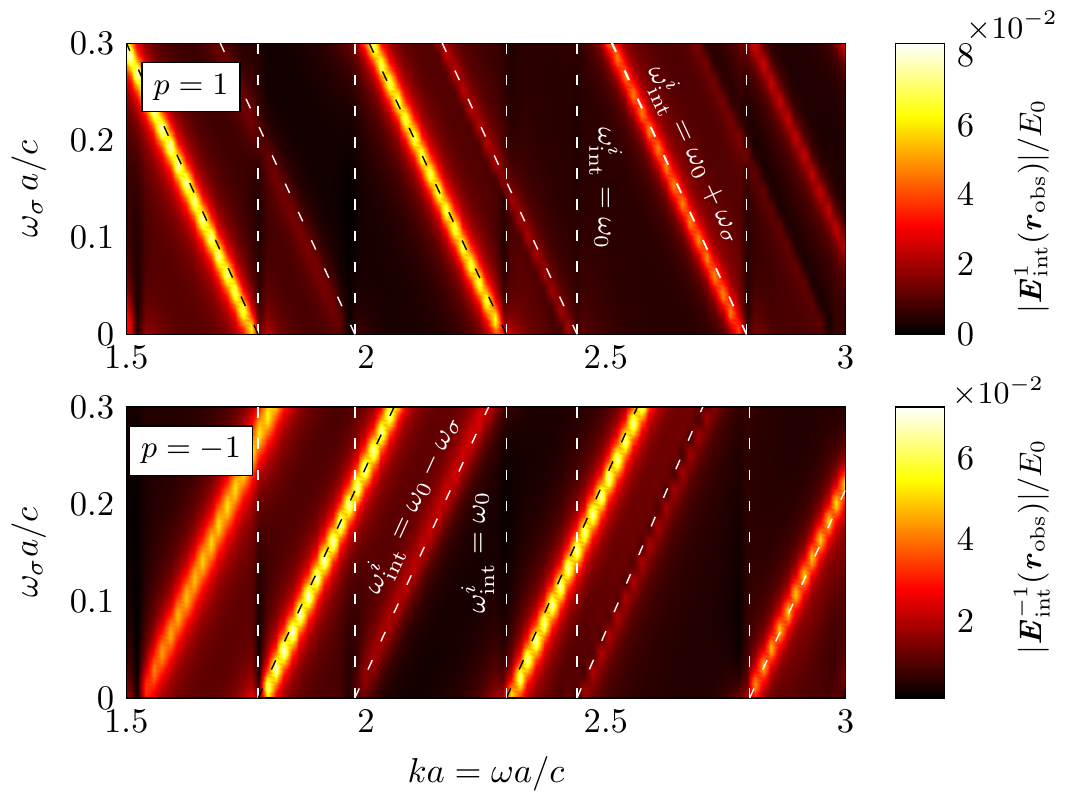}
    \caption{Internal electric field magnitude in the $p=1$ (top) and $p=-1$ (bottom) harmonics from a sphere with relative core permittivity $\varepsilon_\T{r}=6.25$ and time-varying shell resistance given by \eqref{eq:r-single-harm} with $r_0 /\eta = 10^{-2}$ and $\gamma = 0.25$.}
    \label{fig:detuning-resonant}
\end{figure}

Similarly, in the resonant coupling regime we expect that internal field magnitudes will be maximized for the harmonic frequency $\omega_p$ when that frequency corresponds to one of the internal resonances $\{\omega_\T{int}^i\}$ of the PEC-coated dielectric core, i.e., when
\begin{equation}
    \omega_\T{int}^i = \omega_p = \omega_0 + p\omega_\sigma
    \label{eq:internal-resonance}
\end{equation}
is satisfied.  In Fig.~\ref{fig:detuning-resonant}, we plot the normalized field magnitude sampled at $\V{r}_\T{obs} = -\V{\hat{z}}a/2$ in the first  upconverted ($p=1$) harmonic and the first downconverted ($p=-1$) harmonic using the same setup as in Fig.~\ref{fig:detuning-weak}, with the exception that now the resistance is assigned as $r_0 / \eta = 10^{-2}$ and the diagonal dashed lines correspond to the condition prescribed in \eqref{eq:internal-resonance}. For both the $p=1$ and the $p=-1$ cases, it is clear that maximization of internal fields is governed by the condition described in \eqref{eq:internal-resonance}, with little dependence on the electrical size at the excitation frequency. The opposing slopes of the \mbox{$p=-1$} case further show that maximization of the internal field magnitudes also occurs when the downconverted harmonic corresponds to the internal resonance of the system. 


\section{Conclusions}

In this work, we develop a quasi-analytic solution to scattering problems involving spheres coated with time-varying conductive or resistive shells.  Implementation of the method requires only the evaluation of special functions and the inversion of relatively small matrices. This simplicity allows for rapid exploration of high-dimensional parameterizations, as demonstrated by the selected examples.  Though the problem being studied is simple in nature, it represents a class of likely realizations of time-varying structures, i.e., time-varying surface impedances implemented through the use of densely patterned time-varying lumped components \cite{wu2019serrodyne,taravati2020space}, modulation of inherently surface-based materials (e.g., carbon nanostructures \cite{salary2018electrically}), or modulation of plasma conductivity \cite{singletary2021}.  Should it be feasible to construct such a structure through the use of switches or other elements, the results from the examples presented here indicate that several interesting behaviors, particularly up- and down-conversion in radiation and near-fields, could be engineered through careful choice of system parameters.  Additional behaviors may be uncovered through further exploration of multi-sphere systems using the proposed T-matrix formulation.

Several assumptions limit the proposed method to simple homogeneous media and we hypothesize that some of these assumptions may be lifted by more general treatments of the problem.  Specifically, the use of inhomogeneous or anisotropic conductive shells, inclusion of layered media, adaptation to time-varying reactive surfaces or impedance boundary conditions, and combination with methods derived for time-varying dielectrics all stand as tasks for future work in this area.  We expect that the study of these generalizations will be aided by a combination of the field, transmission line, and T-matrix formulations presented in this work. The assumption that all time scales in the problem are much greater than the relaxation time of the material permits the time-varying material to be approximated as having an instantaneous, nondispersive response.  This non-physical approximation may be accurate in low-frequency systems and simplifies the overall formulation presented in this work, but it may not be suitable in all scenarios.  For more accurate solution of higher frequency problems, material dispersion may be accounted for by incorporating methods developed for the analysis of time-varying dispersive lumped elements \cite{jayathurathnage2021time}. 








\appendices

\section{Special function notation}
\label{sec:special-functions}
Throughout this work we adopt the notation of \cite{losenicky2020method} to describe spherical vector harmonics.  Here we clarify two critical points regarding this notation.

The radial dependence of spherical harmonics' tangential field components are defined via the functions $\T{R}_{\tau\ell}^{(\rbkind)}$, where $\tau$ and $\ell$ are two components of the superindex $\alpha = \{\tau\sigma m\ell\}$.  For purposes of this work, we require only harmonics with values of $\tau = 1,~2$.  These functions are given by 
\begin{equation}
    \T{R}_{1\ell}^{(\rbkind)} = \T{Z}_\ell^{(\rbkind)}(kr) = \frac{\hat{\T{Z}}_\ell^{(\rbkind)}(kr)}{kr}
\end{equation}
\begin{equation}
    \T{R}_{2\ell}^{(\rbkind)} = \frac{(kr\T{Z}_\ell^{(\rbkind)}(kr))'}{kr} = \frac{\hat{\T{Z}}_\ell^{(\rbkind)\prime}(kr)}{kr}
\end{equation}
where $\T{Z}_\ell^{(\rbkind)}$ denotes spherical Bessel and Hankel functions of varying kinds, most notably $\T{Z}_\ell^{(1)} = \T{J}_\ell$ and $\T{Z}_\ell^{(4)} = \T{H}^{(2)}_\ell$.  The use of $\hat{~}$ denotes the Riccati-Bessel and Riccati-Hankel form of these functions. 
The use of dual superindices $\alpha$ and $\bar{\alpha}$ indicates a toggling of the value of $\tau$ between 1 and 2, i.e.,
\begin{equation}
    \alpha = {1\sigma m \ell}\quad \rightarrow \quad \bar{\alpha} = {2\sigma m \ell} 
\end{equation}
and
\begin{equation}
    \alpha = {2\sigma m \ell}\quad \rightarrow \quad \bar{\alpha} = {1\sigma m \ell} .
\end{equation}
As suggested in by the appearance of these forms in expansions for electric and magnetic fields in Sec.~\ref{sec:tmat}, the choice of $\tau$ corresponds to a given spherical harmonic $\V{u}_\alpha^{(\rbkind)}$ being of transverse electric (TE) or transverse magnetic (TM) type.

\section{Reconciling T-matrix and transmission line approaches}
\label{sec:app-tl-and-tmat}
Here we outline the procedure for establishing equivalence between the transition matrix and transmission line approaches of Secs.~\ref{sec:tmat} and \ref{sec:tl}. 
A change of variables in \eqref{eq:tmat-def} from modal coefficients to modal voltages relates the transition and reflection matrices via
\begin{equation}
    \M{T} = \M{R}_{\alpha,4}^{-1}\boldsymbol{\Gamma}\M{R}_{\alpha,1}.
\end{equation}
Rearranging the above expression and substituting \eqref{eq:t-mat-final} yields
\begin{equation}
    \M{\Gamma} 
    = \left[\M{R}_{\alpha,1\T{d}}\left(\Xmat_\alpha+\Ymat_\alpha\right)^{-1}\M{R}_{\alpha,1}^{-1}-\M{1}\right].
    \label{eq:app-tl-gamma}
\end{equation}
From the definition of the wave admittances in \eqref{eq:admittance-mat}, we have
\begin{equation}
    \M{Y}_{\alpha,4} - \M{Y}_{\alpha,1\T{d}} = -\T{j}\gamma_\alpha \eta^{-1}\M{W}_\alpha\M{R}_{\alpha,4}^{-1}\Xmat_\alpha\M{R}_{\alpha,1\T{d}}^{-1}.
\end{equation}
Additionally, from the definition of the operator $\Ymat_\alpha$ in \eqref{eq:ymatrix-long} we may write
\begin{equation}
    \Ymat_\alpha = -\T{j}\eta \gamma_\alpha\M{W}_\alpha^{-1}\M{R}_{\alpha,4}\boldsymbol{\sigma}\M{R}_{\alpha,1\T{d}}.
\end{equation}
From the above expressions, we may construct the first term on the right-hand side of \eqref{eq:app-tl-gamma} as
\begin{multline}
    \M{R}_{\alpha,1\T{d}}\left(\Xmat_\alpha+\Ymat_\alpha\right)^{-1}\M{R}_{\alpha,1}^{-1} \\= -\T{j}\gamma_\alpha \eta^{-1}\M{W}_\alpha(\M{Y}_{\alpha,4}-\M{Y}_\T{in})^{-1}\M{R}_{\alpha,4}^{-1}\M{R}_{\alpha,1}^{-1}
\end{multline}
Substituting this into the definition of $\M{\Gamma}$ and rearranging gives
\begin{multline}
    \M{\Gamma} = 
    (\M{Y}_{\alpha,4}-\M{Y}_\T{in})^{-1}\\\cdot
    \left(-\T{j}\gamma_\alpha \eta^{-1}\M{W}_\alpha \M{R}_{\alpha,4}^{-1}\M{R}_{\alpha,1}^{-1} - \M{Y}_{\alpha,4}+\M{Y}_\T{in}\right)
    \label{eq:app-gamma-2}
\end{multline}
Expanding two of the central terms above using the definition of the operator $\mathcal{W}_\alpha$ in \eqref{eq:w-alpha}, several cancellations lead to
\begin{equation}
    -\T{j}\gamma_\alpha \eta^{-1} \M{W}_\alpha\M{R}_{\alpha,4}^{-1}\M{R}_{\alpha,1}^{-1} - \M{Y}_{\alpha,4} = -\M{Y}_{\alpha,1},
\end{equation}
which, when substituted into \eqref{eq:app-gamma-2} yields the expected form of the reflection matrix in \eqref{eq:gamma-mat-final}.

\bibliographystyle{IEEEtran}
\bibliography{main}

\begin{IEEEbiography}[{\includegraphics[width=1in,height=1.25in,clip,keepaspectratio]{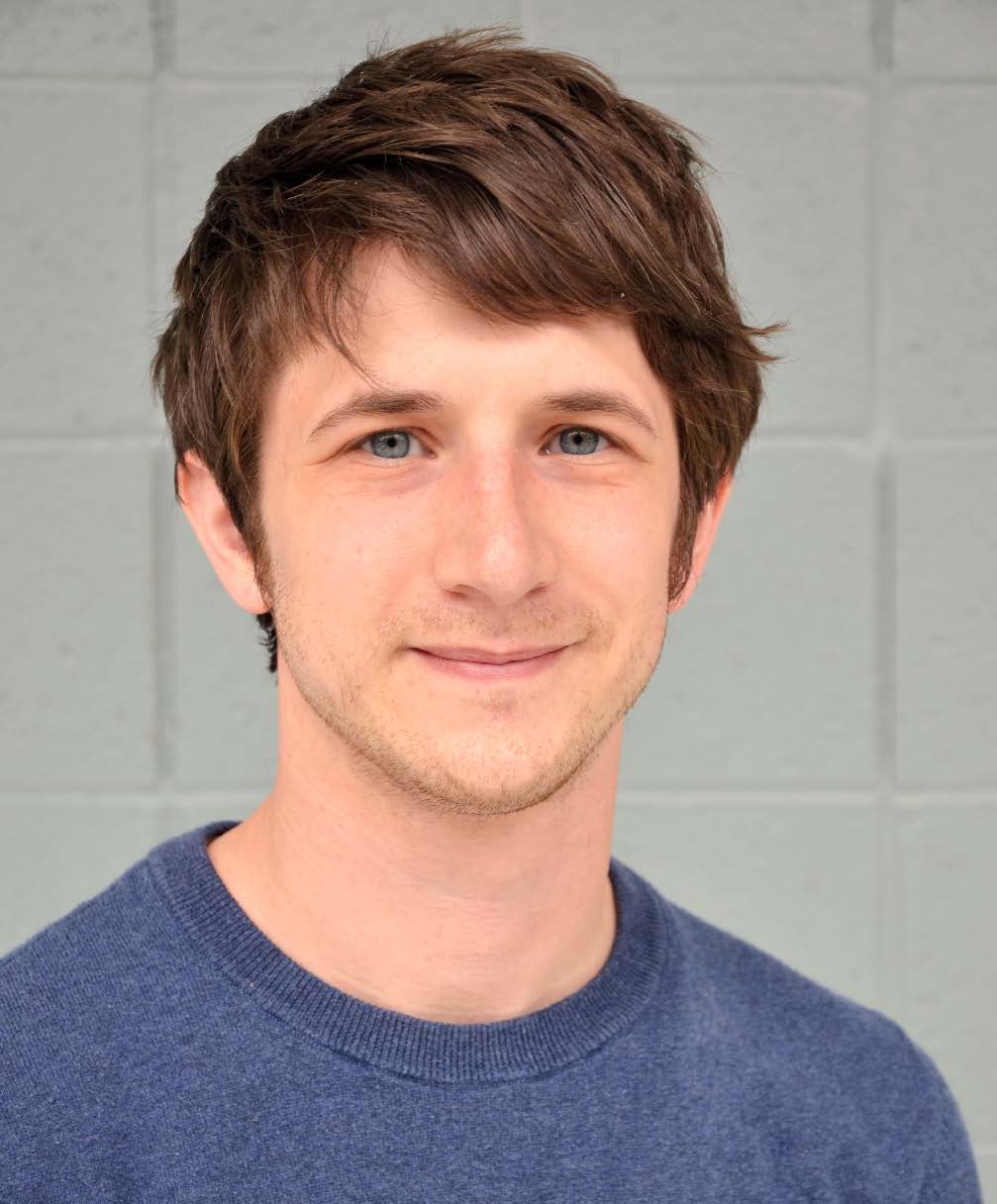}}]{Kurt Schab}
(S'09, M'16) is an Assistant Professor of Electrical Engineering at Santa Clara University, Santa Clara, CA USA. He received the B.S. degree in electrical engineering and physics from Portland State University in 2011 and the M.S. and Ph.D. degrees in electrical engineering from the University of Illinois at Urbana-Champaign in 2013 and 2016, respectively.  From 2016 to 2018 he was a Postdoctoral Research Scholar at North Carolina State University in Raleigh, North Carolina.  His research focuses on the intersection of numerical methods, electromagnetic theory, and antenna design.  
\end{IEEEbiography}

\begin{IEEEbiography}[{\includegraphics[width=1in,height=1.25in,clip,keepaspectratio]{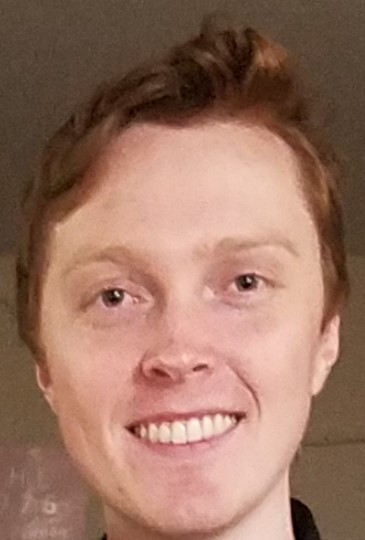}}]{Bradley Shirley} received the B.S. degree in electrical engineering in 2020 and the M.S. degree in electrical and computer engineering in 2021 from Santa Clara University, Santa Clara, CA USA. He is currently an RF electrical engineer at SLAC National Accelerator Laboratory in the RF Accelerator Research division working in the Advanced RF Systems department. 
\end{IEEEbiography}

\begin{IEEEbiography}[{\includegraphics[width=1in,height=1.25in,clip,keepaspectratio]{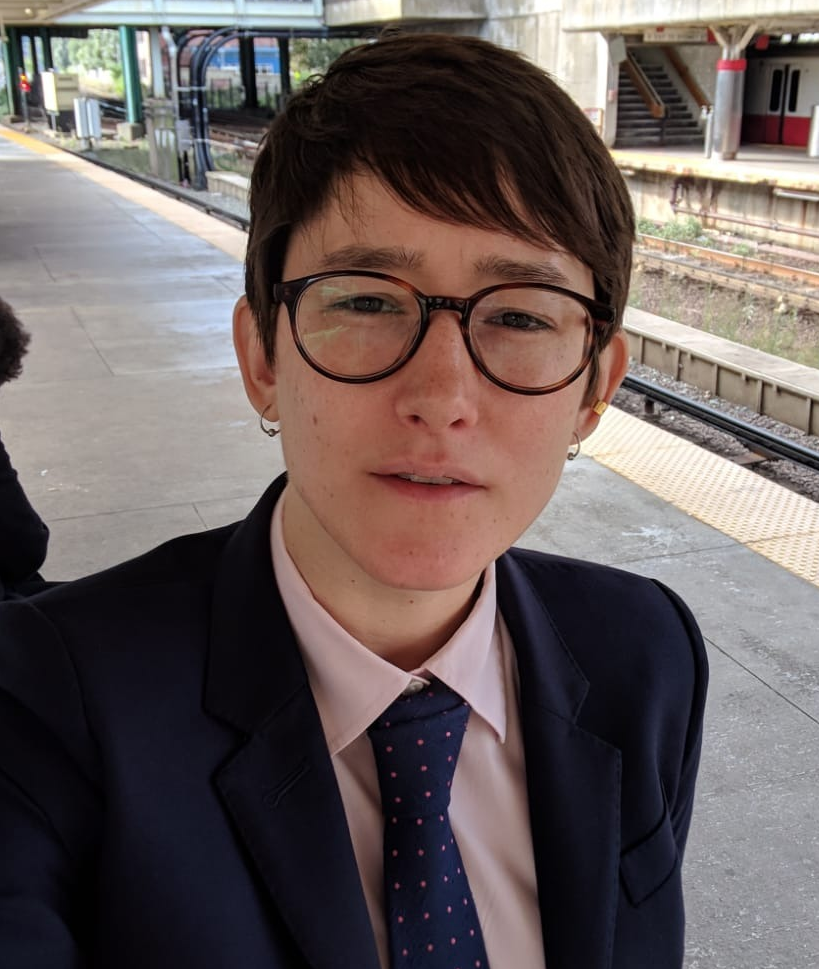}}]
{K.C. Kerby-Patel} (S’06-M’09- SM’14) received the B.S., M.S., and Ph.D. degrees in electrical engineering from the University of Illinois at Urbana-Champaign in 2003, 2005, and 2009, respectively. In fall 2014 she joined the Engineering Department at UMass Boston as an Assistant Professor. Prior to joining the faculty at UMass Boston, she was a Lead Communications Engineer in Applied Electromagnetics at the MITRE Corporation. Her research interests are in applied electromagnetics, particularly the intersection of antenna theory and microwave circuit techniques with new electromagnetic problems and applications. She was awarded the DARPA Young Faculty Award in 2015 for her work on low profile antennas with high impedance ground planes.
\end{IEEEbiography}

\end{document}

%% file: figures/fig-01-sphere.tex
    \newcommand\pgfmathsinandcos[3]{%
        \pgfmathsetmacro#1{sin(#3)}%
        \pgfmathsetmacro#2{cos(#3)}%
    }
    \tikzset{%
        >=latex, 
        inner sep=0pt,%
        outer sep=2pt,%
        mark coordinate/.style={inner sep=0pt,outer sep=0pt,minimum size=3pt,
            fill=black,circle}%
    }

    \begin{tikzpicture}

    \def\R{1.5} 
    \def\angEl{35} 
    \def\angleLongitudeP{-110} 
    \def\angleLongitudeQ{-45} 
    \def\angleLatitudeQ{30} 
    \def\angleLongitudeA{-20} 

    \draw[->] (0,0) -- (2,0) node[right]{$z$};
    \shade[ball color=orange!80,opacity=.75,draw=black] (0,0) circle (\R);
    \tdplotsetmaincoords{90+\angEl}{-5}
    \begin{scope}[tdplot_main_coords]
    \path (0,0,0) coordinate (O);
    \draw[fill=blue!50,opacity=0.5] plot[variable=\t,domain=0:90] (xyz spherical cs:radius=\R,longitude=0,latitude=\t) -- (O) -- cycle;
    \draw[fill=blue!30,opacity=0.5] plot[variable=\t,domain=0:90] (xyz spherical cs:radius=\R,longitude=60,latitude=\t) -- (O) -- cycle;
    \draw[fill=blue!10,opacity=0.5] plot[variable=\t,domain=0:60] (xyz spherical cs:radius=\R,longitude=\t,latitude=0) -- (O) -- cycle;
    \end{scope}
    
    \node at (0.5,0.4) {$\varepsilon_\T{d}$};
        \node at (0.4,-0.5) {$\mu_\T{d}$};
    \node at (-0.7,-0.4) {$\sigma(t)$};
    
    \draw (-2.5,-0.5) -- (-2.5,0.5);
    \draw (-2.7,-0.5) -- (-2.7,0.5) node[above] {$e^\T{i}(t)$};
    \draw (-2.9,-0.5) -- (-2.9,0.5);
    \draw[->] (-3.1,0)--(-2.1,0);
    
    \begin{scope}[shift={(5,2)},rotate=30]
        \draw (-2.5,-0.5) -- (-2.5,0.5);
        \draw (-2.7,-0.5) -- (-2.7,0.5) node[above] {$e^\T{s}(t)$~~~~};
        \draw (-2.9,-0.5) -- (-2.9,0.5);
        \draw[->] (-3.1,0)--(-2.1,0);
    \end{scope}

    \end{tikzpicture}    

%% file: equations/g-eq.tex
\begin{figure*}
\setcounter{tempEQCounter}{\value{equation}}
\setcounter{equation}{24}
\begin{multline}
    g_n =  \T{j}^{-1}\mu_\T{r}^{-1}\sqrt{\varepsilon_\T{r}}^{-1}\left[\sqrt{\mu_\T{r}}\Hh_n(ka)\Jh_n'(k_\T{d}a) -\sqrt{\varepsilon_\T{r}}\Hhp_n(ka)\Jh_n(k_\T{d}a)\right]c_n \\+ \omega\sqrt{\varepsilon_\T{r}\mu_\T{r}}^{-1}\Hhp_n(ka)\int_{-\infty}^\infty \omega^{\circ-1} \eta\hat{\sigma}(\omega-\omega^\circ)c_n(\omega^\circ)\Jh_n'(k^\circ_\T{d}a) \T{d}\omega^\circ.
    \label{eq:g-c-tm-long}
\end{multline}
\setcounter{equation}{\value{tempEQCounter}}
\hrulefill
\vspace*{4pt}
\end{figure*}

%% file: equations/b-and-c-matrix-eqs.tex
\begin{figure*}
\setcounter{tempEQCounter}{\value{equation}}
\setcounter{equation}{41}
\begin{multline}
    \M{B}_n = \eta \sqrt{\varepsilon_\T{r}\mu_\T{r}}^{-1} 
    \begin{bmatrix}
    \omega_{-K}\Hhp_n(k^{-K}a) & 0 & 0 & 0 \\
    0 & \omega_{-K+1}\Hhp_n(k^{-K+1}a) & 0 & 0\\
    0 & 0 & \ddots & 0\\
    0 & 0 & 0 & \omega_{K}\Hhp_n(k^{K}a)
    \end{bmatrix}\\
\begin{bmatrix}
    \hat{\sigma}^{0} &  \hat{\sigma}^{-1} & \hdots & \hat{\sigma}^{-2K} \\
    \hat{\sigma}^{1} &  \hat{\sigma}^0 & \hdots & \hat{\sigma}^{-2K+1} \\
    \vdots & \vdots & \vdots & \vdots\\
    \hat{\sigma}^{2K} &  \hat{\sigma}^{2K-1} & \hdots & \hat{\sigma}^0 \\
    \end{bmatrix}
    \begin{bmatrix}
    \omega^{-1}_{-K}\Jh_n'(k^{-K}_\T{d}a) & 0 & 0 & 0 \\
    0 & \omega_{-K+1}^{-1}\Jh_n'(k^{-K+1}_\T{d}a) & 0 & 0\\
    0 & 0 & \ddots & 0\\
    0 & 0 & 0 & \omega_{K}^{-1}\Jh_n'(k^{K}_\T{d}a)
    \end{bmatrix}
    \label{eq:bmatrix-long}
\end{multline}
\setcounter{equation}{\value{tempEQCounter}}
\end{figure*}

\begin{figure*}
\setcounter{tempEQCounter}{\value{equation}}
\setcounter{equation}{44}
\begin{multline}
    \M{D}_n = \eta \sqrt{\varepsilon_\T{r}\mu_\T{r}}^{-1} 
    \begin{bmatrix}
    \omega_{-K}\Hh_n(\omega_{-K} a/c) & 0 & 0 & 0 \\
    0 & \omega_{-K+1}\Hh_n(\omega_{-K+1} a/c) & 0 & 0\\
    0 & 0 & \ddots & 0\\
    0 & 0 & 0 & \omega_{K}\Hh_n(\omega_{K} a/c)
    \end{bmatrix}\\
\begin{bmatrix}
    \hat{\sigma}^{0} &  \hat{\sigma}^{-1} & \hdots & \hat{\sigma}^{-2K} \\
    \hat{\sigma}^{1} &  \hat{\sigma}^0 & \hdots & \hat{\sigma}^{-2K+1} \\
    \vdots & \vdots & \vdots & \vdots\\
    \hat{\sigma}^{2K} &  \hat{\sigma}^{2K-1} & \hdots & \hat{\sigma}^0 \\
    \end{bmatrix}
    \begin{bmatrix}
    \omega_{-K}^{-1}\Jh_n(k^{-K}_\T{d}a) & 0 & 0 & 0 \\
    0 & \omega_{-K+1}^{-1}\Jh_n(k^{-K+1}_\T{d}a) & 0 & 0\\
    0 & 0 & \ddots & 0\\
    0 & 0 & 0 & \omega_{K}^{-1}\Jh_n(k^{K}_\T{d}a)
    \end{bmatrix}
    \label{eq:dmatrix-long}
\end{multline}
\hrulefill
\vspace*{4pt}
\end{figure*}
\setcounter{equation}{\value{tempEQCounter}}

%% file: equations/y-matrix-equation.tex
\begin{figure*}
\setcounter{tempEQCounter}{\value{equation}}
\setcounter{equation}{66}
\begin{multline}
    \Ymat_\alpha = -\T{j}\eta \gamma_\alpha 
    \begin{bmatrix}
    W_\alpha^{-K} & 0 & 0 & 0 \\
    0 & W_\alpha^{-K+1} & 0 & 0\\
    0 & 0 & \ddots & 0\\
    0 & 0 & 0 & W_\alpha^{K}
    \end{bmatrix}^{-1}
    \begin{bmatrix}
    \Ra^{(4)}(k^{-K}a) & 0 & 0 & 0 \\
    0 & \Ra^{(4)}(k^{-K+1}a) & 0 & 0\\
    0 & 0 & \ddots & 0\\
    0 & 0 & 0 & \Ra^{(4)}(k^{K} a)
    \end{bmatrix}\\
    \begin{bmatrix}
    \hat{\sigma}^{0} &  \hat{\sigma}^{-1} & \hdots & \hat{\sigma}^{-2K} \\
    \hat{\sigma}^{1} &  \hat{\sigma}^0 & \hdots & \hat{\sigma}^{-2K+1} \\
    \vdots & \vdots & \vdots & \vdots\\
    \hat{\sigma}^{2K} &  \hat{\sigma}^{2K-1} & \hdots & \hat{\sigma}^0 \\
    \end{bmatrix}
    \begin{bmatrix}
    \Ra^{(1)}(k_\T{d}^{-K}a) & 0 & 0 & 0 \\
    0 & \Ra^{(1)}(k_\T{d}^{-K+1}a) & 0 & 0\\
    0 & 0 & \ddots & 0\\
    0 & 0 & 0 & \Ra^{(1)}(k_\T{d}^{K}a)
    \end{bmatrix}
    \label{eq:ymatrix-long}
\end{multline}
\hrulefill
\vspace*{4pt}
\end{figure*}
\setcounter{equation}{\value{tempEQCounter}}

%% file: figures/fig-02-tline.tex
    \begin{circuitikz}[
		text pos/.store in=\tpos,text pos=0.5,
		text anchor/.store in=\tanchor,text anchor={north:12pt},
		Tline/.style={
			draw,
			postaction={decorate,decoration={markings,mark=at position 10pt with {\coordinate (a) at (90:3.5pt);}}},
			postaction={decorate,decoration={markings,mark=at position \tpos with {\node at (\tanchor){\small #1};}}},
			postaction={decorate,decoration={markings,mark=at position \pgfdecoratedpathlength-10pt with {\coordinate (b) at (-90:3.5pt);\draw[fill=yellow!10](a) rectangle (b);}}}
		},
		]
			\draw[Tline,text anchor=90:12pt](-1,0) -- (2,0);
			\draw[Tline=$\M{Y}_{1/4}$,text anchor=90:12pt](-1,1.5) -- (2,1.5);
			\draw[Tline=$\M{Y}_{1\T{d}}$,text anchor=90:12pt](2,1.5) -- (5,1.5);
			\draw[Tline,text anchor=90:12pt](2,0) -- (5,0);
			\draw (2,0) to[/tikz/circuitikz/bipoles/length=1cm,generic,l_=$~\V{\sigma}$] (2,1.5);
			\draw[dashed] (2,-1) node[below]{$r=a$}-- (2,0);
			\draw[->] (1.25,-0.5) node[below left] {$\M{Y}_\T{in}$} --(1.25,0.75)--(1.5,0.75);
			
			\node[tape, fill=white, draw=none, minimum height=0.7cm, minimum width=0.5cm, rotate=90] at (-0.8,0) {};
			\node[tape, fill=white, draw=none, minimum height=0.7cm, minimum width=0.5cm, rotate=90] at (-0.8,1.5) {};
			\node[tape, fill=white, draw=none, minimum height=0.7cm, minimum width=0.5cm, rotate=90] at (4.8,0) {};
			\node[tape, fill=white, draw=none, minimum height=0.7cm, minimum width=0.5cm, rotate=90] at (4.8,1.5) {};
			\node at (5,0.75) {$\cdots$};
			\node at (-1,0.75) {$\cdots$};
			\end{circuitikz}